\RequirePackage{rotating}

\documentclass[acmsmall,screen,nonacm]{acmart}
\AtBeginDocument{%
  }

\usepackage[ruled,vlined,linesnumbered]{algorithm2e}
\usepackage{subcaption}
\usepackage{amsmath}
\usepackage{multicol}
\usepackage{multirow}
\usepackage{booktabs}
\usepackage{url}

\usepackage[final, commandnameprefix=ifneeded]{changes}
\usepackage{amsfonts}
\usepackage{caption}
\usepackage{xcolor}      %
\usepackage{tikz}        %
\usepackage{pgfplots}   %
\pgfplotsset{compat=1.16}
\usepackage{pdflscape}
\usepackage{rotating}
\usepackage{colortbl}
\usepackage{xcolor}
\usepackage{rotating}
\usepackage{extarrows}

\definecolor{epflcolor}{RGB}{66,146,198}
\definecolor{blifcolor}{RGB}{220,70,70}

\definecolor{dieTwol}{HTML}{B85450}
\definecolor{dieTwof}{HTML}{F8CECC}

\definecolor{dieThreel}{HTML}{6C8EBF}
\definecolor{dieThreef}{HTML}{DAE8FC}

\newcommand{\cadd}[1]{{\color{black}{#1}}} 
\newcommand{\cdel}[1]{}
\newcommand{\xadd}[1]{{\color{black}{#1}}} 

\begin{document}

\title{Interconnect-Aware Logic Resynthesis for Multi-Die FPGAs}

\author{Xiaoke Wang}
\orcid{0009-0003-8435-3478}
\affiliation{
  \institution{Ghent University}
  \city{Ghent}
  \country{Belgium}
}
\email{xiaoke.wang@ugent.be}

\author{Raveena Raikar}
\orcid{0000-0002-7344-2518}
\email{raveena.raikar@ugent.be}

\author{Markus Rein}
\orcid{0009-0003-5490-999X}
\email{markus.rein@ugent.be}
\affiliation{
  \institution{Ghent University}
  \city{Ghent}
  \country{Belgium}
}


\author{Ruiqi Chen }
\orcid{0000-0001-6837-5675}
\affiliation{
  \institution{Vrije Universiteit Brussel}
  \city{Brussels}
  \country{Belgium}
}
\affiliation{
  \institution{Ghent University}
  \city{Ghent}
  \country{Belgium}
}
\email{ruiqi.chen@ugent.be}

\author{Chang Meng}
\orcid{0000-0003-3134-7771}
\affiliation{
  \institution{Eindhoven University of Technology}
  \city{Eindhoven}
  \country{The Netherlands}
}
\email{c.meng@tue.nl}
\authornote{Corresponding author.}

\author{Dirk Stroobandt}

\orcid{0000-0002-4477-5313}
\affiliation{
  \institution{Ghent University}
  \city{Ghent}
  \country{Belgium}
}
\email{dirk.stroobandt@ugent.be}

\renewcommand{\shortauthors}{Wang et al.}

\begin{abstract}
Multi-die FPGAs enable device scaling beyond reticle limits but introduce severe interconnect overhead across die boundaries. Inter-die connections, commonly referred to as super-long lines (SLLs), incur high delay and consume scarce interposer interconnect resources, often dominating critical paths and complicating physical design. 
\cadd{To address this, this work proposes an interconnect-aware logic resynthesis method that restructures the LUT-level netlist to reduce the number of SLLs. The resynthesis engine uses die partitioning information to apply logic resubstitutions,
which simplifies local circuit structures and eliminates SLLs.
By reducing the number of SLLs early in the design flow, 
prior to \cdel{downstream }physical implementation, the proposed method shortens critical paths\cadd{, alleviates pressure on scarce interposer interconnect resources,} and improves overall physical design flexibility.
We further build a tool flow for multi-die FPGAs by integrating the proposed resynthesis method with packing and placement.}
Experimental results \cadd{on the EPFL benchmarks show that, compared with a state-of-the-art framework, the proposed method reduces the number of SLLs by up to 24.8\% for a 2-die FPGA and up to 27.38\% for a 3-die FPGA.}
On MCNC benchmarks, our \deleted{framework}\xadd{tool flow} achieves an average \cadd{SLL }reduction of 1.65\%\cdel{ in inter-die nets and 10.64\% in inter-die edges,}
while preserving \cdel{overall }placement quality\cdel{ and backend stability}.
On Koios benchmarks, where \cdel{the opportunity for SLL elimination is more limited}\cadd{fewer removable SLLs exist}, 
\cdel{selected}\cadd{several} designs still exhibit considerable inter-die edge reductions.
\cadd{Overall, the results confirm}\cdel{highlighting the effectiveness of logic-level interconnect optimization}\cadd{ that reducing inter-die connections at the logic level is an effective approach} for multi-die FPGAs.
\end{abstract}



\received{Day Month Year}
\received[revised]{Day Month Year}
\received[accepted]{Day Month Year}

\maketitle

\section{Introduction}
Interposer-based multi-die integration has emerged as a 
\cdel{scalable 2.5D integration approach}\cadd{practical solution} 
to overcome the \cdel{maximum die area imposed by lithographic reticle size constraints}\cadd{reticle size limits of lithography}, 
\cadd{enabling large systems to be implemented by partitioning designs across multiple dies~\cite{radojcic2017more}.}
\cadd{This approach is now widely adopted in both \textit{application-specific integrated circuits} (\textit{ASICs}) and \textit{field-programmable gate arrays} (\textit{FPGAs})~\cite{graening2025chipletpartcostawarepartitioning25d,Hahn2014fpga}.
However, distributing a design across dies inevitably introduces \textit{inter-die connections},
\textit{i.e.}, links that cross boundaries between different dies,
which are also referred to as \textit{super-long lines} (\textit{SLLs}).}

\cadd{Compared to \textit{intra-die connections} that remain within a single die,
inter-die connections raise two major challenges.
First, they incur significantly larger delay
because the signals need to traverse long physical distances across dies through the interposer and pass through additional interface circuitry. 
Thus, timing paths including inter-die connections are much slower, 
and even a small number of such connections can noticeably increase the critical path delay
This effect is particularly profound in multi-die FPGAs, 
where signals can cross die boundaries only through dedicated inter-die links, making timing closure more difficult to achieve~\cite{boutros2023into,Nasiri2016TVLSI,Hahn2014fpga}.
Second, excessive inter-die connections increase physical design complexity. 
Inter-die connections consume scarce die-to-die links, 
which constrain block placement and tighten routing constraints. 
Prior work on single-die FPGAs has shown that higher interconnect complexity leads to longer physical design runtime~\cite{wang2024slip}. 
In multi-die CAD flows, this effect can be further amplified.
Inter-die connections tightly couple implementations on different dies,
so physical implementation decisions on one die affect others.
This coupling limits independent processing of dies, reduces parallelism in the design flow, and increases overall implementation difficulty and runtime.}
\cadd{To address these issues,
prior work has explored reducing inter-die connections for FPGAs at different stages of the CAD flow.}
These efforts include multi-die-aware placement~\cite{LEAPS2024TCASI} and
\cdel{HLS-level}\cadd{co-optimization of high-level synthesis} directive\cadd{s} \cdel{and}\cadd{with} floorplanning\cdel{ co-optimization}~\cite{guo2021autobridge_fccm,FACO2023FPGA}, 
highlighting that inter-die delay is a critical concern across different abstraction levels.
Among \cdel{these approaches}\cadd{them}, partitioning-based FPGA CAD flows \cdel{aim to }reduce inter-die \cdel{communication}\cadd{connections} through timing-aware assignment of blocks across dies~\cite{kuo2018pin, liao2020pin}. 
Previous work \cdel{has tightly integrated}\cadd{integrates} timing-driven partitioning into the open-source \textit{Verilog-to-Routing} (\texttt{VTR}) framework~\cite{jasonluu2017vtr7}, 
demonstrating improvements in critical-path delay and maximum operating frequency \cdel{(\texttt{$F_{max}$}) }by discouraging cuts on timing-critical paths~\cite{iyer_partitioning-based_2025, raveena2024HEART}.
\cdel{Partitioning-base approaches typically rely on an explicit trade-off between minimizing cut size and maintaining balanced partitions~\cite{raveena2022slip}. Allowing higher imbalance can reduce the number of inter-die connections by clustering strongly connected logic into fewer dies. However, excessive imbalance leads to congested dies with limited physical design flexibility, while other dies remain underutilized.}%

\cadd{While these methods can effectively reduce the inter-die connections, they operate on a netlist that has already been fixed by logic synthesis and technology mapping, 
meaning the underlying logic structure that generates the inter-die connections remains unchanged.}
\cdel{As a result, partitioning and refinement}\cadd{Thus, \replaced{they}{existing methods}} can only \cdel{rearrange existing connectivity}\cadd{redistribute existing connections across dies},
rather than re-evaluating whether inter-die connections are logically necessary.
\cadd{Even with high-quality partitioning, 
many inter-die connections may persist because partitioning does not change the circuit logic, 
so signals required on multiple dies must still be communicated across die boundaries.}
As a result, existing flows across different abstraction levels mitigate inter-die connections, rather than fundamentally reducing reliance on them from the logic level.

\cadd{To overcome this limitation, we explore}
\textit{logic resynthesis} \cdel{enables direct modification of circuit}\cadd{as a means to directly modify the netlist} connectivity\cdel{ by exploiting functional flexibility and don't-care conditions at the Boolean level, particularly for LUT-level combinational logic}\cadd{ at the logic level}.
\cadd{Operating on mapped netlists,
such as the \textit{lookup table} (\textit{LUT})-level netlists commonly used in FPGA design,
logic resynthesis modifies the netlist so that internal signals are computed and shared differently, using
functional equivalences and don't-care conditions.
Conventionally, logic resynthesis has been used to optimize area, delay, or power,
while we propose to leverage it to optimize inter-die connectivity in multi-die FPGAs.}
This logic-level restructuring capability \cdel{provides}\cadd{introduces} a\cdel{n}\cadd{ new} optimization \cdel{opportunity}\cadd{dimension} 
\cadd{for reducing inter-die connections at their logical origin by selectively resynthesizing Boolean logic,}
thereby lowering \cdel{\emph{super-long line}}\cadd{SLL} utilization and mitigating interposer-induced timing \cdel{penalties}\cadd{degradation}.

\cadd{The feasibility of this approach is supported by our benchmark analysis. 
We observe that a large fraction of signal connectivity originates from LUT-level soft logic rather than fixed architectural hard blocks.
This indicates that many inter-die connections arise from flexible logic structures that can be modified, 
rather than rigid block-level interfaces. 
Consequently, even after \cdel{min-cut} partitioning, 
a large fraction of inter-die connections remains amenable to modification through logic-level restructuring.}




\cdel{In this work}\cadd{Building on the above insight}, 
we propose\cdel{ \texttt{ResynMD},} an interconnect-aware logic resynthesis \cdel{framework}\cadd{method} for multi-die FPGAs.
\cadd{Unlike existing approaches that operate at the partitioning or physical design stages,
the proposed method directly eliminates
unnecessary inter-die connections at their logical origin in the LUT-level netlists.
To the best of our knowledge,
this is the first work that optimizes inter-die connections for multi-die FPGAs at the logic resynthesis stage.}
Our main contributions are summarized as follows:
\begin{enumerate} 
\item 
\cadd{We formulate interconnect-aware logic resynthesis as an optimization problem that directly targets the reduction of SLLs in multi-die FPGAs.}
\item 
To solve the optimization problem,
we propose an SLL-aware resynthesis technique \cdel{that}\cadd{called \texttt{ResynMD}.
It} iteratively reconstructs local logic in a LUT-level netlist to eliminate unnecessary inter-die connections,
while preserving functional correctness and \cdel{limiting area overhead}\cadd{avoiding area increase}.
\item 
\cadd{We develop a multi-die FPGA CAD tool flow that integrates the proposed resynthesis technique after partitioning and before physical design, which complements existing \cdel{multi-die FPGA CAD }flows.}
\end{enumerate} 


Experimental results show that\cadd{ on EPFL benchmarks,
compared to a state-of-the-art partitioning-based multi-die FPGA CAD flow,} 
our \cdel{framework}\cadd{method} reduces the number of SLLs by up to 24.8\% for 2-die and 27.38\% for 3-die FPGAs.
\cadd{On MCNC benchmarks, 
our method achieves an average reduction of 3.24\% in inter-die connections, 
while preserving overall placement quality.}
On \cdel{heterogeneous }Koios benchmarks,
the proposed method can also \cdel{improve}\cadd{reduce} inter-die \cdel{connectivity}\cadd{connections} while preserving \cadd{post-}placement quality\cdel{ and slightly improving runtime and bounding box cost}.

\cadd{The remainder of this paper is organized as follows.
Section~\ref{sec:pre} describes the preliminaries.
Section~\ref{sec:met} elaborates the proposed interconnect-aware logic resynthesis method for multi-die FPGAs.
The experimental results are presented in Section~\ref{sec:exp}, 
followed by the conclusion and future work in Section~\ref{sec:con}.}

\section{Preliminaries}\label{sec:pre}

This section reviews the architectural and algorithmic \cdel{foundations relevant to }\cadd{background on} interconnect-aware logic resynthesis for multi-die FPGAs. 
We first \cdel{introduce}\cadd{describe} the characteristics of multi-die FPGA fabrics and the resulting non-uniform interconnect landscape, highlighting the role of inter-die \cdel{links}\cadd{connections} and their impact on timing and \cdel{routability}\cadd{physical design}. 
We then summarize partitioning-based \cadd{multi-die FPGA} CAD flows.\cdel{ commonly used in multi-die implementations, 
along with key metrics for evaluating partition quality and placement cost.} 
Finally, we \cadd{introduce the LUT-level network representation.}

\begin{figure}[!htbp]
  \centering
  \includegraphics[width=1.0\linewidth]{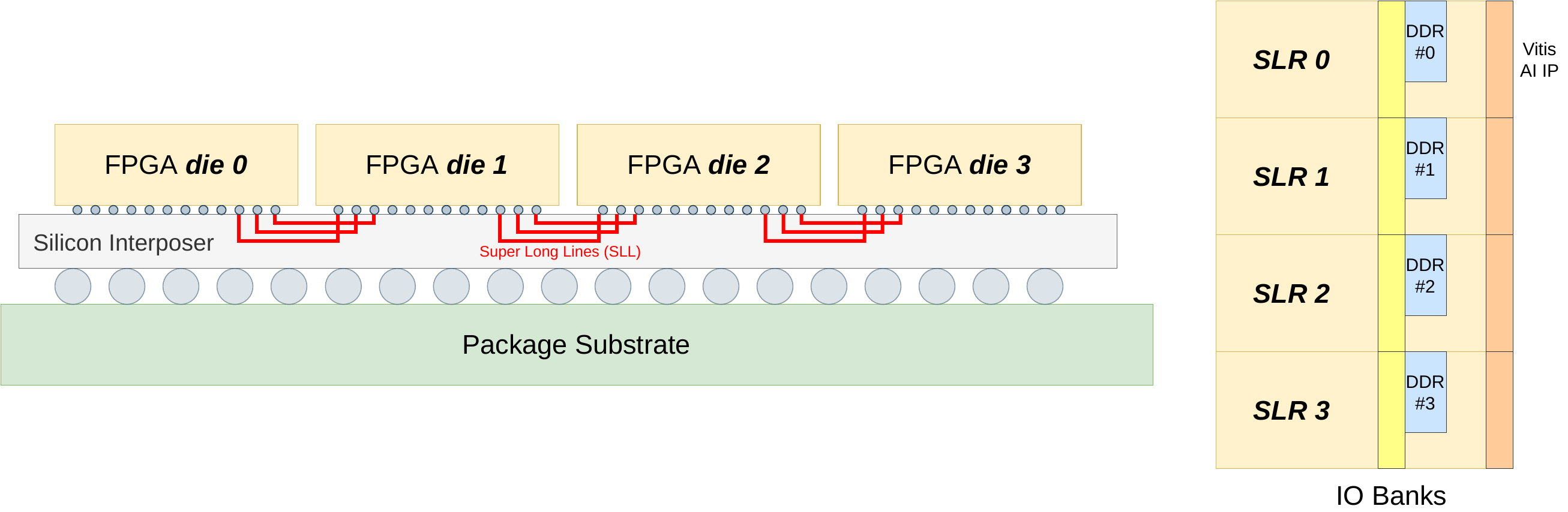}
  \caption{Lateral and top views of a four-die FPGA system (AMD Xilinx Alveo U250~\cite{amd-u200-ds962}).}
  \label{fig:lateral_md}
  \Description{}
\end{figure}

\subsection{Multi-Die FPGAs}

Multi-die FPGAs integrate multiple FPGA dies within a single package, typically using a silicon interposer to provide inter-die communication. This 2.5D integration paradigm enables device scaling beyond reticle limits, but results in non-uniform interconnect characteristics between intra-die and inter-die routing resources. Signals that cross die boundaries traverse interposer links, commonly modeled as \cdel{\emph{super-long lines}}\cadd{SLLs}, as illustrated in Figure~\ref{fig:lateral_md}.

\begin{figure*}[!htbp]
  \centering

  \begin{minipage}[c]{0.30\textwidth}
    \centering
    \includegraphics[width=\linewidth]{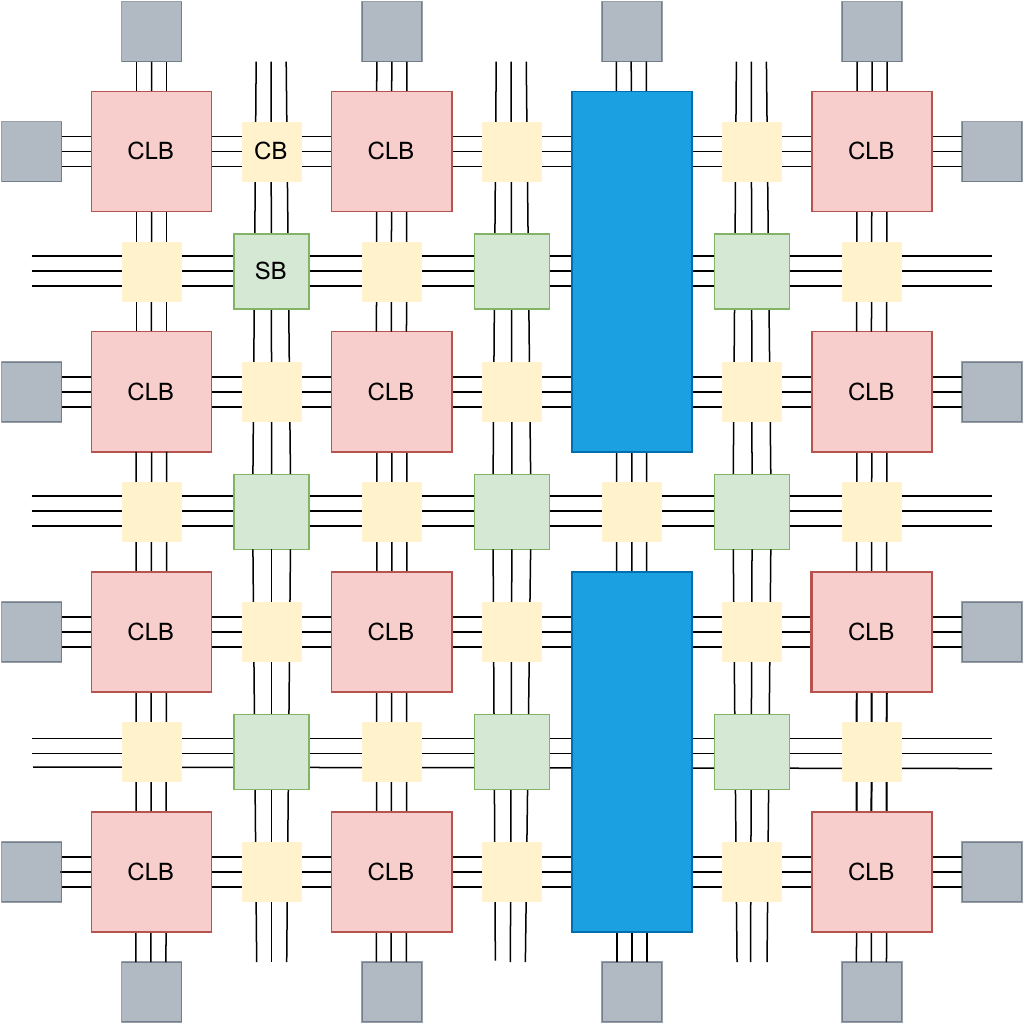}
    \subcaption{}
    \label{fig:md_fabric_a}
  \end{minipage}\hfill
  \begin{minipage}[c]{0.33\textwidth}
    \centering
    \includegraphics[width=\linewidth]{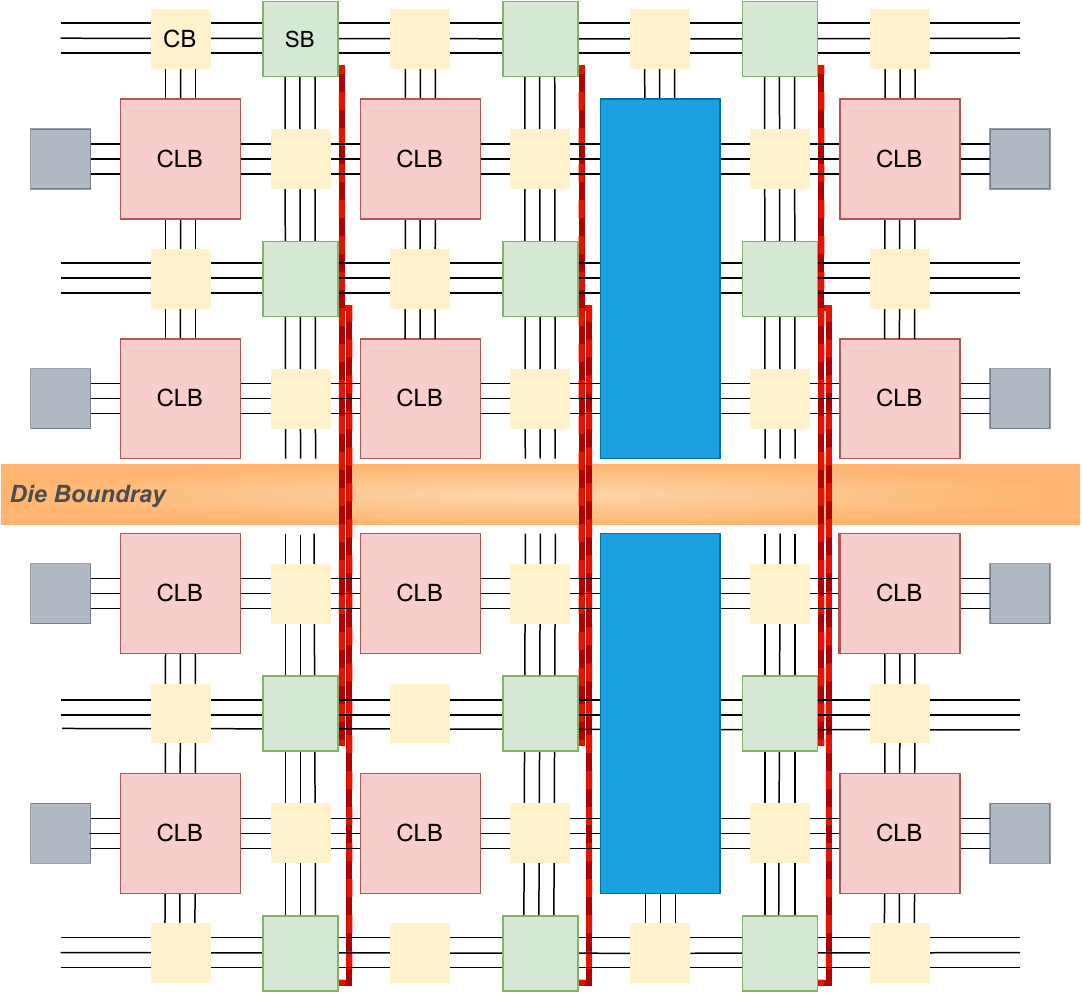}
    \subcaption{}
    \label{fig:md_fabric_b}
  \end{minipage}\hfill
  \begin{minipage}[c]{0.33\textwidth}
    \centering
    \includegraphics[width=\linewidth]{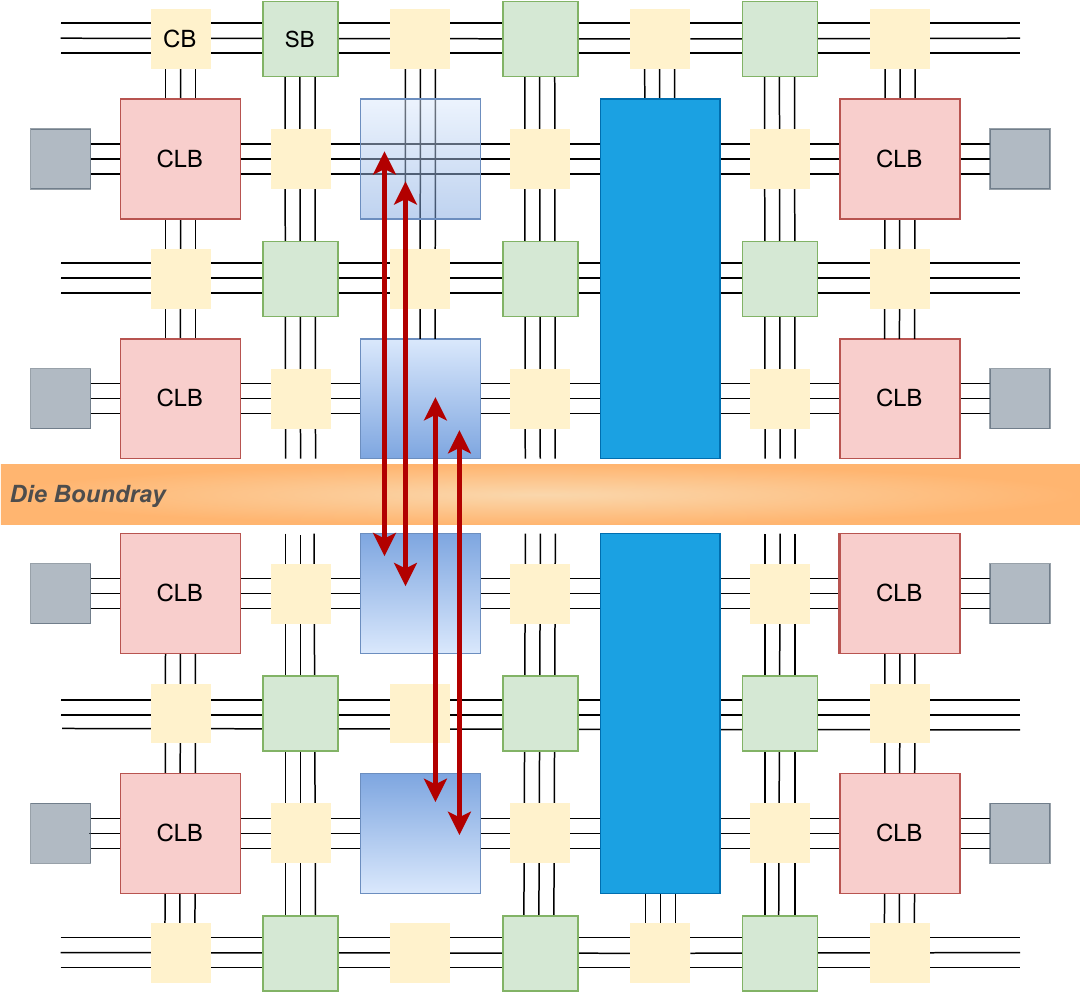}
    \subcaption{}
    \label{fig:md_fabric_c}
  \end{minipage}

  \caption{Fabric-level interconnect abstractions for single-die and multi-die FPGAs:
  (a) a conventional single-die fabric;
  (b) a multi-die fabric with routing-based interposer interfaces;
  and (c) a multi-die fabric with tile-based interposer interfaces.}
  \label{fig:md_fabric}
  \Description{}
\end{figure*}

At the fabric level, this non-uniformity is reflected in the top-view architecture. As shown in Figure~\ref{fig:md_fabric_a}, a single-die FPGA provides a largely homogeneous routing fabric. In contrast, the multi-die fabric in Figure~\ref{fig:md_fabric_b} exposes inter-die communication through a limited number of fixed-location, fixed-length SLLs connected to the routing fabric via dedicated switch boxes. Compared to intra-die routing wires, SLLs offer reduced routing flexibility and distinct delay characteristics.

Commercial multi-die FPGAs may further employ tile-based interposer interfaces with dedicated interface tiles (e.g., Laguna columns)~\cite{ravishankar2018sat} shown in Figure~\ref{fig:md_fabric_c}. Since the logic resynthesis techniques studied in this work operate on a generic logic netlist representation, they are agnostic to the specific implementation of the interposer. For clarity and compatibility with open-source academic CAD tools, we adopt the routing-based abstraction throughout this paper.

Table~\ref{tab:wire_delay} summarizes representative delay values for intra-die routing wires and inter-die interposer links used in this work. 
\cadd{The percentage indicates the fraction of routing tracks of each wire type within the total routing channel width.}
These values are derived from published technology-scaled delay models~\cite{Oleg2016Delay,Sun2019Pattern} and serve as parameters for characterizing the timing differences between routing resources.

\begin{table}[!htbp]
\centering
\caption{Intra-die (22\,nm) and inter-die (45\,nm interposer) wire delays.}
\label{tab:wire_delay}
\begin{tabular}{lccc}
\hline
\textbf{Wire Type} & \textbf{Technology} & \textbf{Delay (ps)} & \textbf{Percentage (\%)} \\
\hline
L1  & 22\,nm\cadd{ (Intra-die)}  & 76.2   & 50 \\
L2  & 22\,nm\cadd{ (Intra-die)}  & 94.9   & 25 \\
L6  & 22\,nm\cadd{ (Intra-die)}  & 212.0  & 25 \\
L36 & 45\,nm (Inter-die) & 2223.7 & 48 \\
\hline
\end{tabular}
\end{table}

\subsection{Partitioning-Based Multi-Die FPGA Flow}
\begin{figure}[!htbp]
    \centering

    \begin{subfigure}[t]{1\linewidth}
        \centering
        \includegraphics[width=.6\linewidth]{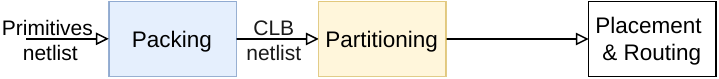}
        \caption{Packing-first flow.}
        \label{fig:strategy_a}
    \end{subfigure}

    \vspace{0.3em}

    \begin{subfigure}[t]{1\linewidth}
        \centering
        \includegraphics[
          width=.6\linewidth
        ]{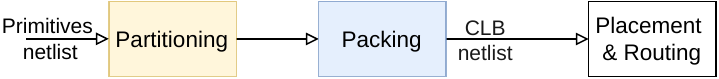}
        \caption{Partitioning-first flow.}
        \label{fig:strategy_b}
    \end{subfigure}
    \caption{Two partitioning-based CAD strategies for multi-die FPGAs:
    (a) a packing-first flow, where packing precedes \cdel{die }partitioning; and
    (b) a partitioning-first flow, where the partitioning is performed before packing.
    }
    \label{fig:packing_order}
    \Description{}
\end{figure}

By decomposing a monolithic design into multiple \cdel{die-level }sub-circuits\cadd{ on multiple dies}, 
partitioning reduces the \cdel{effective }optimization space\cadd{ of the CAD problems} and enables scalable and partially parallel CAD processing, 
while still requiring careful coordination of inter-die \cdel{connectivity}\cadd{connections}.
Figure~\ref{fig:packing_order} illustrates two representative partitioning-based CAD strategies for multi-die FPGA implementation, which differ in whether partitioning is performed before or after packing. 
In packing-first flows, 
\cdel{logic is}\cadd{LUTs are} first clustered into architecture-specific blocks\cdel{ (e.g., CLBs or LABs)}\cadd{, such as configurable logic blocks (CLBs)}, 
and partitioning is subsequently applied to these packed blocks~\cite{iyer_partitioning-based_2025, graening2025chipletpartcostawarepartitioning25d}. 
This ordering closely reflects physical constraints and simplifies downstream placement and routing, but fixes the logic connectivity at the granularity of packed blocks. 
In contrast, partitioning-first flows \cdel{apply partitioning at the primitive network level}\cadd{partition the primitive netlist} before packing, 
allowing logic connectivity to be considered before architectural constraints are fully imposed\cadd{.}
Several state-of-art multi-die CAD flow\cadd{s}~\cite{iyer_partitioning-based_2025, graening2025chipletpartcostawarepartitioning25d, raveena2022slip, raveena2024HEART} use the partitioning-first paradigm.
The netlist is first partitioned into \cdel{die-level }sub-circuits\cadd{ on multiple dies}, 
after which packing\cdel{ and}\cadd{,} placement\cadd{, and routing} are performed \cdel{largely independently }for each die. 
\cdel{This “partitioning-first with optimization” paradigm enables subsequent fine-grained optimizations, including logic-level (LUT) resynthesis, instead of prematurely committing to coarse-grained structural decisions.}%

\cadd{Our interconnect-aware resynthesis technique is better suited to the partition-first flow. In this flow, we insert logic resynthesis after partitioning and before packing. Operating on the LUT-level netlist at this stage preserves more freedom for modification because LUTs have not yet been grouped into CLBs. Compared with a CLB-level netlist, the LUT-level representation is finer-grained and exposes more connections, allowing resynthesis to restructure logic and reduce inter-die connections more effectively.}




\xadd{}

\deleted{
\cadd{Unlike partitioning-based frameworks 
that reduce inter-die connections by reorganizing the physical distribution of logic while treating the circuit structure as fixed,
\texttt{ResynMD} directly modifies the logic circuit structure to reduce unnecessary inter-die connections.}
Specifically, \cdel{it}\cadd{\texttt{ResynMD}} operates on a \cdel{partitioned, LUT-level netlist with a fixed die assignment}\cadd{post-mapped LUT-level netlist that has been partitioned across multiple dies. Each LUT is associated with a fixed die assignment determined by the partitioning stage.
Under this constraint,} 
\cdel{and reduces inter-die connectivity through localized logic resynthesis,}%
\cadd{\texttt{ResynMD} applies localized structural transformations to reduce inter-die connections while preserving functional correctness and} 
without altering the partitioning solution or the downstream physical design algorithms.}


\subsection{\cdel{Logic Network}\cadd{LUT-Level Network Representation}}
FPGA logic synthesis typically operates on circuits represented as graph-based logic networks, which provide a structural abstraction of Boolean functionality. 
After \cdel{logic optimization on}\cadd{optimizing} the logic \cdel{graphs}\cadd{networks}, 
technology mapping transforms the circuit into a \cdel{network}\cadd{netlist} of $k$-input \cdel{\textit{lookup tables} (\textit{LUTs})}\cadd{LUTs}\cdel{, 
which constitutes the primitive-level representation used in subsequent CAD stages}. 
The \cdel{resulting }LUT\cdel{ network}\cadd{-level netlist} explicitly defines the logic connectivity exposed to physical design processes, and therefore serves as the interface between logic synthesis and architecture-aware optimization. 
A LUT-level \cdel{network}\cadd{netlist} can be \cdel{defined}\cadd{viewed} as a directed acyclic graph\cdel{ (DAG)} $G = (V, E)$. 
In this graph, each node $v \in V$ represents a $k$-input LUT implementing a local Boolean function $f_v$. 
Each directed edge $(u, v) \in E$ denotes a signal connection from the \cdel{output of }$u$ to \cdel{an input of }$v$.
Operating at the LUT level enables fine-grained optimization of inter-die connectivity by directly manipulating inter-LUT signals while ensuring seamless integration with downstream CAD flows.

For a node $v$, its \textit{transitive fanin} (\textit{TFI}) is the set of nodes that can reach $v$ through directed paths in $G$. 
Similarly, its \textit{transitive fanout} (\textit{TFO}) is the set of nodes reachable from $v$. 
These structural regions characterize the upstream and downstream influence of $v$ in the logic network.
The \textit{maximum fanout-free cone} (\textit{MFFC}) of \cdel{a }node $v$, denoted $\mathrm{MFFC}(v)$, is the maximal subgraph rooted at $v$ such that every node in the subgraph has no fanout outside the subgraph. 
Removing $v$ will disconnect all nodes in $\mathrm{MFFC}(v)$ from the primary outputs. 
The MFFC corresponds to the portion of logic exclusively controlled by $v$, 
and \cdel{replacing}\cadd{removing} $v$ eliminates its entire MFFC.
\section{\cdel{\texttt{ResynMD}: }Interconnect-Aware Resynthesis \cdel{Framework }for Multi-Die FPGAs}\label{sec:met}

This section presents \texttt{ResynMD}, 
an interconnect-aware logic \underline{resyn}thesis method for \underline{m}ulti-\underline{d}ie FPGAs.
We first introduce the motivation and overall tool flow in Section~\ref{subsec:overview}.
Then, Section~\ref{subsec:obj} formulates the SLL-aware resynthesis problem
and optimization objective.
Section~\ref{subsec:components} describes the core components of the proposed resynthesis engine. 
Finally, Section~\ref{subsec:tradeoff} discusses the trade-off between partition imbalance and inter-die connection reduction.

\subsection{Motivation and Overview}\label{subsec:overview}


Modern applications mapped onto multi-die FPGAs typically contain large amounts of soft logic (\textit{i.e.}, logic implemented using LUTs) with \cdel{dense}\cadd{many} interconnections.
\cdel{Under partitioning-based multi-die CAD flows,}
\cadd{Compared with hard blocks such as DSPs and RAMs, 
soft logic is highly flexible in placement and routing
and is therefore more likely to be distributed across dies.
While hard blocks generally reside at fixed sites and are rarely split across dies, LUT-level logic in large designs is frequently partitioned, producing numerous cross-die connections implemented using SLLs.}
Because SLLs incur high delay and limited routing flexibility, 
they frequently dominate critical paths and contribute significantly to routing congestion.

This work focuses on reducing the number of SLLs by targeting the logic-level origins of inter-die connectivity.
These \cadd{inter}-die \cdel{dependencies}\cadd{connections} are not purely \cdel{dictated}\cadd{required} by architectural constraints, 
but often stem from the \cdel{structural organization}\cadd{logic implementation} of the \cadd{mapped} LUT-level \cdel{logic network}\cadd{netlist}.
\cdel{By selectively restructuring logic}\cadd{Consequently, modifying the logic implementation} 
\cadd{can often replace cross-die signals with functionally equivalent in-die implementations, 
thereby reducing the inter-die connections.}
\cadd{By directly restructuring the logic to reduce inter-die connections,
\texttt{ResynMD} tool flow complements existing partitioning-first CAD flows for multi-die FPGAs~\cite{raveena2024HEART,LEAPS2024TCASI} (see Figure~\ref{fig:strategy_b}), 
which optimize inter-die connections through partitioning or placement
without altering the circuit at the logic level.}

\begin{figure}[!htbp]
    \centering
    \includegraphics[width=\linewidth]{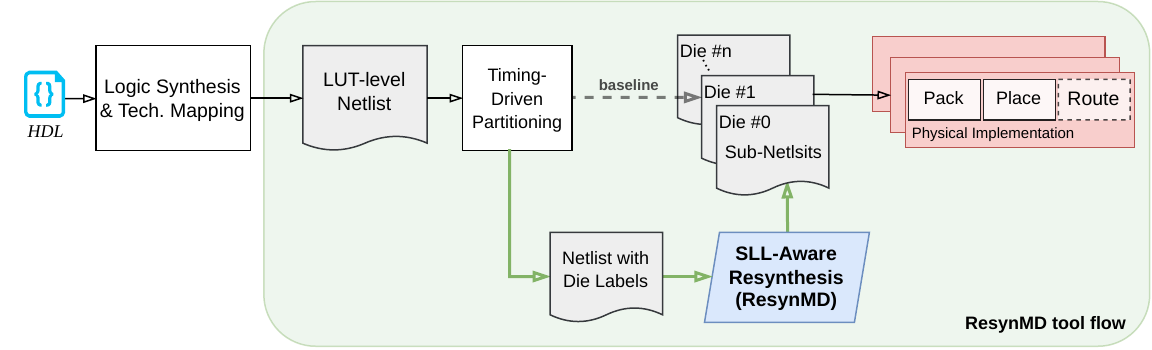}
    \caption{The proposed \texttt{ResynMD} tool flow.
    Compared with the baseline \cdel{\texttt{LiquidMD}}\cadd{partitioning-first CAD} flow\cadd{s (Fig.~\ref{fig:strategy_b})}, 
    this \cdel{CAD }tool flow adds an \cdel{interconnect}\cadd{SLL}-aware resynthesis step between partitioning and \cdel{parallel }physical implementation.}
    \label{fig:flowchart}
    \Description{}
\end{figure}

Figure~\ref{fig:flowchart} shows the \cadd{\texttt{ResynMD} }tool flow.
\cdel{The flow}\cadd{It} starts from a synthesizable \cdel{Verilog design}\cadd{HDL description} and outputs a physical implementation for a multi-die FPGA.
\cadd{It first}
performs \cadd{conventional }logic synthesis and technology mapping \cdel{using Yosys~\cite{yosys2025yosys} and ABC, producing a primitive-level BLIF~\cite{berkeley1992berkeley} netlist as the input to the subsequent partitioning stage}\cadd{to obtain a LUT-level netlist}.
The \cdel{circuit}\cadd{netlist} is \cdel{first}\cadd{then} partitioned \cdel{using an hMETIS-based~\cite{karypis1998hypergraph}, timing-aware partitioning scheme, followed by the proposed interconnect-aware resynthesis stage and the subsequent physical design steps, including packing, placement, and routing.}\cadd{across multiple dies using a timing-aware partitioner,
which reduces inter-die connections on critical paths while balancing logic resources across dies.}
After partitioning, \cdel{the circuit is represented as a single, unified BLIF netlist, wherein each logic node is annotated with a fixed die assignment.}\cadd{each LUT is assigned a fixed die label,
resulting in a partition-annotated netlist in which}
\cdel{At this stage, }inter-die connections \cdel{are explicitly represented as}\cadd{correspond to} nets \cdel{with fanins and fanouts }spanning multiple dies.
\cadd{\texttt{ResynMD}, which is an SLL-aware logic resynthesis step,} is then applied to this partition-annotated \cdel{BLIF }netlist\cdel{ prior to decomposition}.
\cadd{Its objective is to reduce the number of SLLs by restructuring netlist logic while preserving the fixed die assignments.}
Following resynthesis, 
the optimized netlist is \cdel{divided}\cadd{split} into \cdel{independent die-level BLIF}\cadd{per-die sub-}netlists\cdel{, each corresponding to a single die}.
Thereafter, the\cadd{ standard} physical design stages\cdel{—}\cadd{,} including packing, placement, and routing\cdel{—}\cadd{,} are executed \cdel{largely independently }\cdel{and in parallel }for each die\cdel{ with required synchronization}.


Compared to the \cdel{baseline}\cadd{existing partitioning-first CAD flows, 
such as \texttt{LiquidMD}~\cite{raveena2024HEART, Dries2017LiquidFPT, Dries2018MultiPart}}\cdel{ flow},
\cadd{the key distinction of the proposed tool flow in Figure~\ref{fig:flowchart} is the SLL-aware logic resynthesis step that is applied after partitioning but before physical implementation,
as illustrated by} the blue segment \cdel{as illustrated }in Figure~\ref{fig:flowchart}. 
\cadd{To the best of our knowledge, 
\texttt{ResynMD} is the first approach that explicitly targets logic-level restructuring for reducing inter-die connections.
This introduces a new optimization dimension that is complementary to existing partitioning-based and placement-based multi-die optimization techniques and can provide synergistic benefits when integrated into current multi-die FPGA design flows.
The following subsection details the problem formulation and the core components of the proposed SLL-aware resynthesis method.}



\subsection{\cadd{Problem Formulation of SLL-Aware Logic Resynthesis}\cdel{Optimization Target}}\label{subsec:obj}

\cadd{Conventional logic synthesis methods, 
such as ~\cite{mishchenko2006scalable,lee2021simulation,amaru2018improvements,liu2024cbtune},
optimize metrics such as area or delay without considering the multi-die context
and do not explicitly reduce inter-die connections.
In contrast,}
\cdel{This step}\cadd{the proposed SLL-aware logic resynthesis} operates on \cdel{the partitioned, mapped}\cadd{a partitioned} LUT-level \cdel{network}\cadd{netlist}
and\cadd{ directly targets} \cdel{aims to reduce unnecessary} inter-die \cdel{dependencies prior to physical design}\cadd{connections}\cdel{ while preserving functional correctness}. 
\cadd{In this setting,
the die assignment of each LUT is fixed by the partitioning stage, 
and only functionally equivalent logic transformations are permitted.}

Let $G = (V, E)$ denote the \cdel{mapped }LUT-level \cdel{logic network after partitioning}\cadd{netlist with partition labels},
where each node $v \in V$ is assigned to a \cdel{target }die $D(v)$.
\cadd{A connection between nodes $u$ and $v$ is inter-die
(and thus implemented using an SLL)
if $D(u) \neq D(v)$.}
We define the \cadd{\textit{total number of SLLs} (also called \textit{inter-die connectivity})} as
\begin{equation}
N_{sll}(G) = \sum_{(u,v)\in E} \mathbb{I}[\mathcal{D}(u) \neq \mathcal{D}(v)],
\end{equation}
where $\mathbb{I}[\cdot]$ is an indicator function that equals 1 if the \cdel{net}\cadd{connection between $u$ and $v$} crosses the die boundary and 0 otherwise.

\cadd{Our objective is to construct a functionally equivalent netlist $G' \equiv G$
that minimizes the number of SLLs without increasing the netlist area.
Formally, the objective is
\begin{equation}\label{eq:obj_target}
\min_{\substack{G' \equiv G}} N_{sll}(G')
\text{ subject to } Area(G') \leq Area(G),
\end{equation}%
where the function $Area(\cdot)$ measures the netlist area in terms of LUT count.}


\cadd{Because global optimizing Equation~\eqref{eq:obj_target} over all equivalent netlists needs exploring a huge combinatorial search space,}
we adopt a greedy heuristic that \cdel{incrementally}\cadd{iteratively} applies local resubstitution\cadd{~\cite{mishchenko2006scalable}} transformations\cadd{,
which preserve functionality while modifying the fanin structure of nodes.
This approach}
\cdel{which }offers good scalability and computational efficiency for large-scale circuits.
\cdel{At}\cadd{In} each \cdel{step}\cadd{iteration}, 
\cdel{only transformations that}\cadd{a transformation is accepted only if it} strictly reduce\cadd{s the total number of SLLs} and satisf\cdel{y}\cadd{ies the} area constraint\cdel{s are accepted}.
\cadd{More details on the resubstitution procedure are presented in Section~\ref{subsec:components}.}

\subsection{\cadd{Details of SLL-Aware Logic Resynthsis}}\label{subsec:components}

\subsubsection{Greedy Algorithm for SLL-Aware LUT-Level Resynthesis}

\begin{figure}[!htbp]
    \centering
    \begin{subfigure}[t]{0.7\textwidth}
        \centering
        \includegraphics[width=\linewidth]{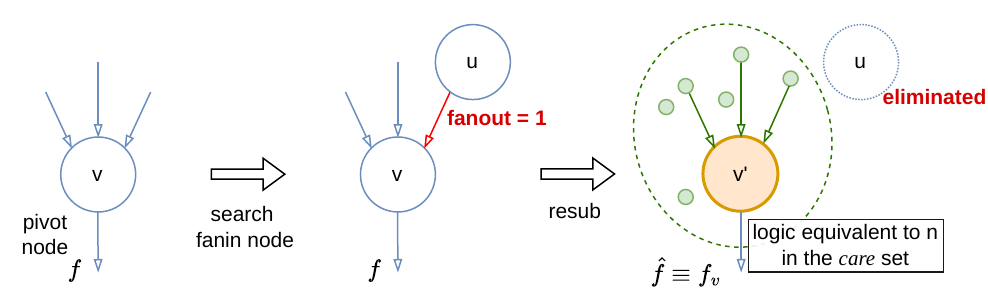}
        \caption{Conventional MFS-base\xadd{d} resubstitution.}
        \label{fig:resub_a}
    \end{subfigure}\\[0.5cm]
    \begin{subfigure}[t]{\textwidth}
        \centering
        \includegraphics[width=\linewidth]{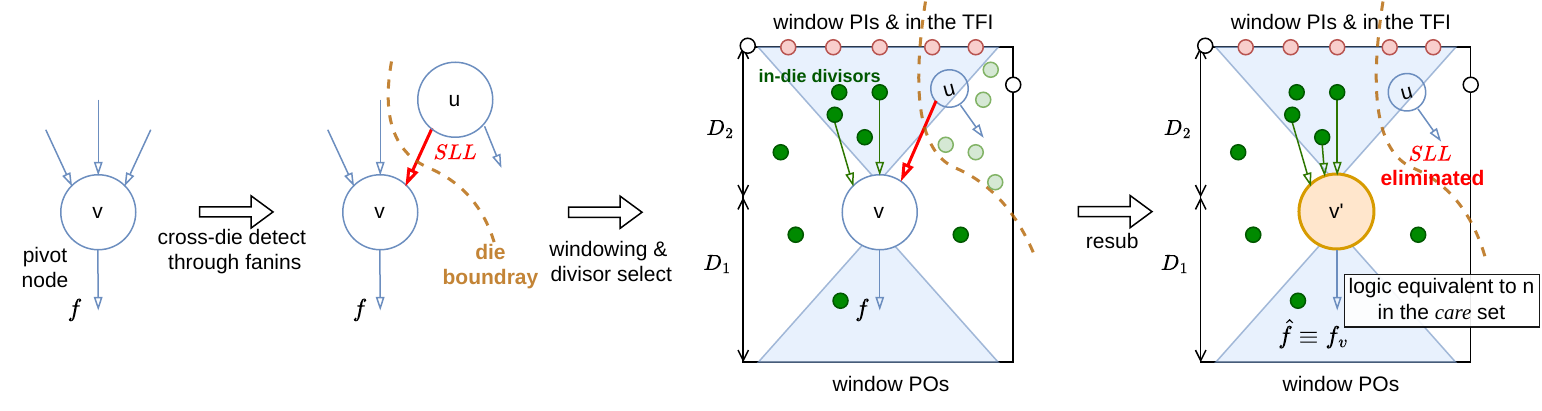}
        \caption{MFS-based interconnect-aware resubstitution.}
        \label{fig:resub_b}
    \end{subfigure}
    \caption{Comparison between conventional MFS-based LUT-level resubstitution
    and the proposed interconnect-aware resubstitution.}
    \label{fig:resub_compare}
    \Description{}
\end{figure}

\begin{algorithm}[!htbp]
\caption{Greedy \cdel{Interconnect-aware}\cadd{SLL-Aware Logic Resynthesis by} LUT-level Resubstitution}
\label{alg:md_resub_greedy}
\small
\KwIn{LUT\cdel{ network}\cadd{-level netlist} $G=(V,E)$, die mapping \cdel{$\mathcal{D}: V\rightarrow\{0,\dots,K-1\}$}\cadd{$\mathcal{D}(v)$ for each node $v\in V$}}
\KwOut{\cdel{Resynthesized network}\cadd{Optimized LUT-level netlist }$G'$ with \cdel{fewer inter-die fanins}\cadd{reduced number of SLLs.}}

Initialize\cadd{ working netlist} $G' \leftarrow G$\;\label{alg:resub:init}

\For{\cadd{each pivot node $v$ in $G'$ in topological order}}{\label{alg:resub:main_loop}
    \If{$\forall u \in \text{fanin}(v): \mathcal{D}(u)=\mathcal{D}(v)$}{\label{alg:resub:skip_same_die_st}
        continue;\tcp*[f]{Skip the node without \cadd{cross}-die fanins}\label{alg:resub:skip_same_die_end}
    }

    \cdel{Build local window $W(v)$}\cadd{Sub-circuit $W(v) \leftarrow \textit{BuildWindow(G', v)}$};\tcp*[f]{Extract local circuit surrounding $v$}\label{alg:resub:build_window}



    \cadd{Divisor set $\mathcal{S}_d(v) \leftarrow \textit{GetSLLAwareDivisors}(W(v), v)$};\tcp*[f]{Get divisors used to re-express $v$}\label{alg:resub:get_divisors}
    
    \cadd{New function $\hat{f}_v \leftarrow \textit{FindEquivFunc}(W(v), v, \mathcal{S}_d(v))$};\tcp*[f]{Find $v$'s equivalent function}\label{alg:resub:find_equiv_func}

    \If{$\hat{f}_v$ exists}{\label{alg:resub:check_equiv_func}
        \cadd{New node $v' \leftarrow LUTImplement(\hat{f}_v)$}\;
        
        \If{%
        \cadd{$\#$cross-die fanins of $v' < \#$cross-die fanins of $v$}\\\label{alg:resub:check_crossdie}
        \cadd{and $Area(v') \le Area(v)$}}{\label{alg:resub:check_area}

            \cadd{$G' \leftarrow \textit{ApplyResubstitution}(G', v, v')$};\tcp*[f]{Replace $v$ with new node $v'$}\label{alg:resub:commit_substitution}
        }
    }
}
\Return{$G'$}\;\label{alg:resub:return}
\end{algorithm}

\cadd{This subsection presents the core algorithm of the proposed SLL-aware logic resynthesis method.
The method builds upon the resynthesis technique \texttt{MFS}~\cite{alanmi2011mfs}.
Unlike conventional \texttt{MFS}, 
which optimizes traditional metrics such as area or delay, 
our new contributions lie in adapting the resynthesis approach to the multi-die context with a specific focus on reducing inter-die connections,
as illustrated in Figure~\ref{fig:resub_compare}.
The inter-die connections are reduced by performing \textit{LUT-level resubstitution}.
Here, \textit{resubstitution}~\cite{alanmi2011mfs, mishchenko2006using} refers to a local logic transformation that replaces a node with a functionally equivalent implementation expressed in terms of a selected set of other nodes (called \textit{divisors}) in the netlist.}

\cadd{As shown in Algorithm 1,
the input of our resynthesis procedure is a partition-annotated LUT-level netlist $G$,
where each node is assigned a die label determined during partitioning.
The output is an optimized netlist $G'$ with fewer inter-die connections, 
which can be directly fed into the subsequent physical implementation stages.
The algorithm initializes the working netlist $G'$ as the input netlist $G$ (Line~\ref{alg:resub:init}).
Then, it processes each node $v$ in $G'$ in topological order (Line~\ref{alg:resub:main_loop}),
ensuring that fanin nodes are handled before their fanouts and preventing cyclic dependencies during resubstitution.
We call the node $v$ being processed the \textit{pivot node}.
For each pivot node, 
the algorithm first checks whether all fanins of $v$ reside on the same die as $v$ (Line~\ref{alg:resub:skip_same_die_st}).
If so, node $v$ has no cross-die fanins and cannot contribute to SLL reduction, 
and is therefore skipped (Line~\ref{alg:resub:skip_same_die_end}).
For node $v$ with cross-die fanins, 
a local \textit{window} $W(v)$ is constructed (Line~\ref{alg:resub:build_window}).
This window is a sub-circuit surrounding node $v$ that
captures the local logic context of $v$ while limiting the scope of optimization to ensure computational tractability.
Within this window,
a set of SLL-aware divisors $\mathcal S_d(v)$ is collected (Line~\ref{alg:resub:get_divisors}), 
which are potential candidate signals used to construct $v$'s new implementation.
These divisors consist only of nodes located on the same die as $v$, 
ensuring that any new implementation of $v$ can be realized without introducing new cross-die fanins and enabling the elimination of existing ones.
Using these divisors, 
the algorithm attempts to derive an equivalent Boolean function $\hat{f}_v$ for node $v$ (Line~\ref{alg:resub:find_equiv_func}).
Note that such a function may not always exist (see Section~\ref{subsec:sat_resub} for details).
If no such function exists, the node $v$ remains unchanged. 
Otherwise, the function $\hat{f}_v$ is implemented using one or more LUTs to create a new node $v'$ (Line~\ref{alg:resub:check_equiv_func}),
which serves as a candidate resubstitution for $v$.
The candidate resubstitution is accepted only if the new node $v'$ has fewer cross-die fanins than the original node $v$,
and does not increase area in terms of LUT count (Lines~\ref{alg:resub:check_crossdie} and~\ref{alg:resub:check_area}).
When both conditions are satisfied, the resubstitution is committed to the working netlist $G'$ (Line~\ref{alg:resub:commit_substitution}), and the algorithm proceeds to the next node.
After all nodes have been processed,
the optimized netlist $G'$ is returned (Line~\ref{alg:resub:return}).}

The above greedy strategy guarantees monotonic reduction of cross-die fanins of each node $v$ in the LUT-level netlist while maintaining scalability for large designs.
Since each accepted resubstitution removes at least one inter-die fanin edge of node $v$ (Line~\ref{alg:resub:check_crossdie}), 
the \cdel{global inter-die connectivity}\cadd{total number of SLLs} $N_{sll}$ decreases monotonically\cdel{.
This ensures} \cadd{ensuring the} convergence of the procedure\cdel{ without requiring global re-optimization}.
\cadd{Moreover, the area constraint in Equation~\eqref{eq:obj_target} is enforced through the condition check in Line~\ref{alg:resub:check_area},
which disallows any resubstitution that increases LUT count.}



\cadd{Algorithm~\ref{alg:md_resub_greedy} relies on three key procedures corresponding to three main components of the SLL-aware resynthesis method:
$\textit{BuildWindow}$, which constructs the local optimization region,
$\textit{GetSLLAware}\textit{Divisors}$, which selects same-die candidate signals, 
and $\textit{FindEquivFunc}$, which constructs an equivalent function, typically via SAT-based resubstitution.
These components are described in detail in Section~\ref{subsec:window}--\ref{subsec:sat_resub}.}

\subsubsection{Window\cdel{ing}\cadd{ Construction}}\label{subsec:window}

\cadd{This step corresponds to Line~\ref{alg:resub:build_window} in Algorithm~\ref{alg:md_resub_greedy}.}
\cadd{We adopt the same window construction strategy as \texttt{MFS}~\cite{alanmi2011mfs}.}
\cadd{A}s illustrated in \cadd{the middle of }Figure~\ref{fig:resub_b}\cadd{,
for each pivot node $v$ (\textit{i.e.}, node being processed for resubstitution),
its window $W(v)$ is a sub-circuit surrounding $v$ that captures the local logic context relevant for $v$'s resubstitution.}
\cdel{For each target node, a window}\cadd{$W(v)$} is formed by expanding its \cdel{transitive fanout (TFO)}\cadd{TFOs} up to a limited depth \textit{$D_1$}. 
The\cadd{n,} $\textit{$D_2$}$ levels\cadd{ of} \cdel{transitive fanin (TFI)}\cadd{TFIs} of the pivot node \cdel{\textit{n}}\cadd{$v$} determines the window \cdel{supports (inputs)}\cadd{\textit{primary inputs (PIs)}}\cdel{, while}\cadd{. After that,} the internal nodes are obtained through a topologically ordered traversal of the induced subgraph\cdel{.}\cadd{, forming the window $W(v)$.}

\cdel{Within each optimization window, 
a set of divisors is identified to define the candidate signal\cdel{ source}s for resubstitution. 
Divisors correspond to logic nodes whose Boolean functions are already available within the window and can be reused to synthesize an alternative implementation of the target node. 
Conceptually, divisors serve as a restricted library of intermediate signals, 
allowing the optimizer to express the target function as a Boolean combination of existing logic rather than introducing new primary inputs. 
To balance optimization quality and complexity, 
divisors are primarily collected from the transitive fanin of the target node and selectively expanded through local fanout relationships, yielding a bounded yet functionally rich candidate set.}%


\subsubsection{Divisor Selection}\label{subsec:divisor}
\xadd{This step corresponds to the computation of the divisor set $\mathcal{S}_d(v)$ in Line~\ref{alg:resub:get_divisors} of Algorithm~\ref{alg:md_resub_greedy}, where candidate divisors are extracted from the local window $W(v)$ and filtered according to their die assignment.}
Divisors are signals within the local window $W(v)$ that can be used to re-express the Boolean function of the pivot node $v$. 
\cdel{They are selected following the conventional \texttt{MFS} methodology, as}\cadd{Their selection process is} illustrated in the middle part of Figure~\ref{fig:resub_b}.
\xadd{When cross-die fanins of the pivot are detected, a local window is constructed around the pivot node.} Window PIs are \deleted{partitioned}\xadd{split} according to whether they belong to the transitive fanin of the pivot. The nodes on structural paths between the pivot and its TFI-related window PIs are collected as potential divisors, excluding the pivot itself and the nodes in its maximum fanout-free cone (MFFC). Additionally, window nodes whose structural support does not depend on non-TFI window PIs are also considered. To limit complexity, divisors \xadd{that} exceed a predefined logic-level bound are discarded, and the total number of divisors is capped.

To incorporate inter-die interconnection awareness, 
each divisor is classified according to its die assignment relative to the \deleted{target} \xadd{pivot} node. 
The divisors located on the same die (\cadd{highlighted in }dark green\cadd{ in Figure~\ref{fig:resub_b}}) as the target are greedily selected \xadd{and constitute the final in-die divisor set $S_d(v)$}.
\deleted{This die-aware divisor selection is illustrated in Figure~\ref{fig:resub_b}.}\

In conventional \cdel{MFS-based }resubstitution, candidate divisors are selected solely based on functional\cdel{ feasibility}\cadd{ or structural criteria without considering their physical die assignment},
\cdel{potentially preserving or even}\cadd{sometimes} introducing \cadd{new }inter-die \cdel{dependencies}\cadd{connections}.
In contrast, the proposed approach enforces a greedy, 
\cdel{die}\cadd{SLL}-aware divisor selection strategy that deterministically  \cdel{prioritizes}\cadd{considers} same-die signals\cdel{ before invoking the SAT-based resubstitution check}. 
This enables the following SAT-based resubstitution \cadd{approach} to eliminate \cdel{avoidable}\cadd{unnecessary} inter-die connections.


\subsubsection{SAT-Based Resubstitution}\label{subsec:sat_resub}

This step corresponds to the Line~\ref{alg:resub:find_equiv_func} of Algorithm~\ref{alg:md_resub_greedy}. 
The goal is to \cdel{get an equivalent}\cadd{compute a} function\cdel{ $\hat{f} \equiv f_v$}\cadd{ $\hat f$ that is equivalent to $f_v$ (the function of the pivot node $v$),} 
\cadd{expressed using a}
subset of in-die divisors $S_d' \subseteq S_d(v)$. The key idea is to selectively eliminate cross-die fanins of the pivot node $v$ by expressing its function $f_v$ using only in-die signals.
For a pivot node $v$, we construct its equivalent function $\hat{f}$ in a structured manner,  as summarized in Algorithm~\ref{alg:algorithm2}. Rather than searching for arbitrary divisor combinations, we restrict the construction to eliminate cross-die dependencies.
\begin{algorithm}[ht]
\caption{\textit{FindEquivFunc}$(W(v), v, \mathcal{S}_d(v))$}
\label{alg:algorithm2}
\small
\KwIn{Local window $W(v)$, pivot node $v$, divisor set $\mathcal{S}_d(v)$}
\KwOut{Equivalent function $\hat{f}$ or $\emptyset$\cadd{ (\textit{i.e.}, $\hat f$ does not exist)}}

Select a cross-die fanin $u \in \text{fanin}(v)$\;\label{alg2:line:select_u}

$BaseDivisors \leftarrow \text{fanin}(v) \setminus \{u\}$\;\label{alg2:line:base_def}

$C \leftarrow ExtractCareSet(W(v), v)$

\If{$\textit{SATBasedExistCheck}(f_v, C, S_d')$}{\label{alg2:line:direct_sat}
    $ S_d' \leftarrow BaseDivisors$\;
    \Return{$\hat{f} \cdel{=}\cadd{\leftarrow} \textit{Interpolate}(S_d')$}\;
}

\For{each divisor $d \in \mathcal{S}_d(v)$}{\label{alg2:line:loop_start}
    $S_d' \leftarrow BaseDivisors \cup \{d\}$\;\label{alg2:line:update_sd}
    \If{$\textit{SATBasedExistCheck}(f_v, C, S_d')$}{\label{alg2:line:sat_check}
        \Return{$\hat{f} \cdel{=}\cadd{\leftarrow} \textit{Interpolate}(S_d')$}\;\label{alg2:line:return_interp}
    }
}

\Return{$\emptyset$}\;\label{alg2:line:return_empty}

\end{algorithm}

\xadd{First, this procedure is triggered by identifying a cross-die fanin 
$u \in \mathrm{fanin}(v)$ and attempting to remove it. 
The base divisor subset $S_d'= BaseDivisors$ is initialized by excluding $u$ while retaining all remaining fanins (Line~\ref{alg2:line:base_def}). If $f_v$ can be expressed using this reduced set under the care constraint, the cross-die edge $(u, v)$ is directly eliminated.}

\xadd{If this direct elimination fails, we traverse the in-die divisor set $S_d(v)$ (Line~\ref{alg2:line:loop_start}). For each candidate divisor $d$, we construct a new divisor subset 
$S_d'$ by augmenting the base divisors with $d$, i.e., by adding a single 
in-die divisor to compensate for the removed cross-die dependency. 
Feasibility is then re-evaluated for this augmented set.}



\xadd{For each constructed divisors subset $S_d'$, a SAT-based existence check~\cite{alanmi2011mfs} is performed under the extracted care set $C$ to determine whether $f_v$ can be equivalently expressed using signals in $S_d'$.  If the SAT-based existence check succeeds, 
an interpolation is extracted to construct the equivalent function $\hat{f}_v$~\cite{lee2007interpolation}. 
The interpolation depends only on signals in $S_d'$  and preserves equivalence under the care set $C$.}
The derived function $\hat{f}_v$ is then \cdel{materialized}\cadd{implemented} as a new LUT node $v'$ whose fanins correspond to the selected divisors in $S_d'$.  Node $v'$ is inserted into the original network, and all fanout connections of the original pivot node $v$ are redirected to $v'$. Subsequently, node $v$ and any newly unreachable logic in its $\mathrm{MFFC}(v)$ are removed from the network. This network update corresponds to line~\ref{alg:resub:commit_substitution} in Algorithm~\ref{alg:md_resub_greedy}. 
\subsubsection{Example}\label{subsec:example}
\cdel{For simplicity and clarity, w}\cadd{W}e use a small \cadd{2-input LUT (LUT2 in short) netlist as an example}\cdel{LUT2 network} to illustrate how \cdel{interconnect-aware resubstitution}\cadd{our SLL-aware resynthesis technique} can reduce redundant \cadd{inter}-die \cdel{dependencies under the MFS framework}\cadd{connections}.
\cadd{Note that LUT2 is used here for ease of presentation, and the proposed method is applicable to general LUT sizes.}

\begin{figure}[!htbp]
    \centering
    \begin{subfigure}[t]{\textwidth}
        \centering
        \includegraphics[width=.6\linewidth]{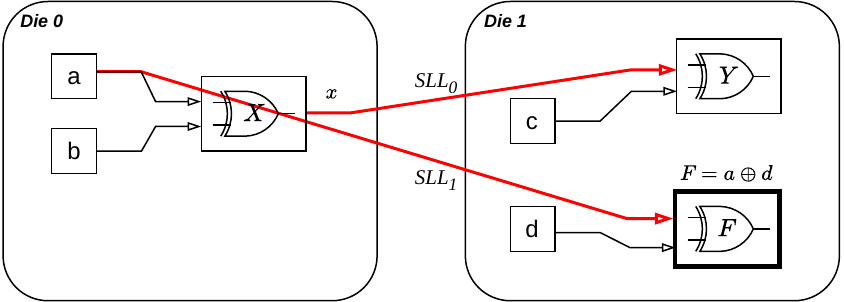}
        \caption{\cadd{Pre-resynthesis: t}he original \cdel{circuit after partitioning}\cadd{LUT-level netlist} before resubstitution ($F$ is pivot node).}
        \label{fig:resub_example_a}
    \end{subfigure}\\[0.5cm]
    \begin{subfigure}[t]{\textwidth}
        \centering
        \includegraphics[width=.6\linewidth]{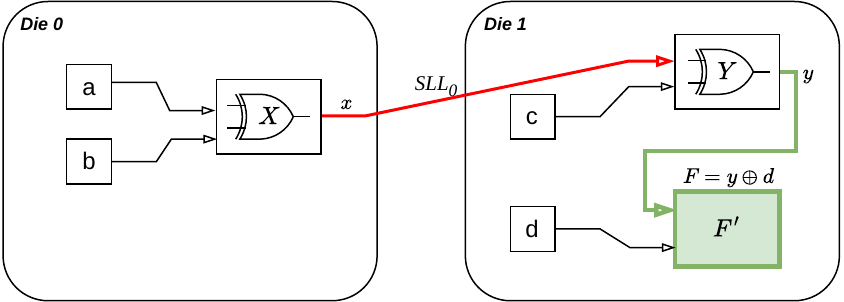}
        \caption{\cdel{Interconnect}\cadd{After resynthesis: SLL}-aware resubstitution eliminating the redundant \cadd{inter}-die dependency on $a$.}
        \label{fig:resub_example_b}
    \end{subfigure}
    \caption{\cdel{Pre-resynthesis and post-resynthesis example.}\cadd{Example of SLL-aware resynthesis.}
    }
    \label{fig:resub_compare_example}
    \Description{}
\end{figure}

\textbf{Pre-resynthesis logic.}
\cadd{As shown in Figure~\ref{fig:resub_example_a}, c}onsider a partitioned LUT\cdel{ network}\cadd{-level netlist} mapped onto a two-die FPGA. Let $a,b,c\cadd{,d}$ be primary inputs.
\cdel{A}\cadd{The} LUT node $X$ is placed on Die~0, 
while nodes $Y$\cdel{, $d$,} and \cdel{the pivot node }$F$ are placed on Die~1.
\cadd{Node $F$ is the pivot node to be resubstituted.}
Signal \cdel{$X$}\cadd{$x$} crosses the \cdel{boundary of the die to the feed node $Y$}\cadd{die boundary}, 
forming an \cadd{inter}-die connection\cadd{ between $X$ on Die~0 and $Y$ on Die~1}.
In addition, the pivot node $F$\cadd{ on Die~1} depends directly on signal $a$\cadd{ on Die~0}, introducing an extra \cadd{inter}-die fan-in that we aim to eliminate.

The logic functions \cadd{in the 2-input LUTs }are defined as:
\begin{equation}
\begin{aligned}
X &= a \oplus b \\
Y &= X \oplus c \\
F &= a \oplus d
\end{aligned}
\label{eq:example}
\end{equation}

\textbf{Don't-care and care set definition.}
\cdel{In MFS-based resubstitution,}\cadd{As discussed in Section~\ref{subsec:sat_resub},} 
the functionality of a pivot node is required to be preserved only on its care set, 
\cdel{while deviations on don't-care minterms are allowed.}\cadd{and the don't-care cases can be exploited to enable more flexible resubstitution.}
For illustrative purposes, we 
\cadd{assume that }input assignments with $b \neq c$ are \cdel{determined to be outside of the care set}\cadd{don't-cares} of the pivot node $F$\cdel{ under the constraints derived from SAT}.
\cadd{In other words, we only care about the functional equivalence under the input cases when $b=c$}.
\cdel{Accordingly, these assignments are treated as \emph{don't-cares (DCs)} at $F$, yielding}\cadd{The don't-care and care conditions are defined as}:
\begin{equation}\label{eq:dc_care}
    \begin{aligned}
    DC(a,b,c,d) &= b \oplus c, \\
    C(a,b,c,d)  &= \lnot (b \oplus c).
    \end{aligned}
\end{equation}


\textbf{\xadd{Resynthesis process and }post-resynthesis logic.}
\xadd{Node $F$ is the pivot node to start the process.}
\xadd{As the current example is simple, following the windowing method introduced in Section~\ref{subsec:window}, all of the LUT nodes across two dies will be covered in the \emph{window}.} \xadd{During the divisor selection step, }within the resynthesis window, 
only the \textit{in-die} signals are available as candidate divisors, including $Y$\cadd{, $c$} and $d$.
\cdel{Using MFS-based resubstitution under the care set $C$}\cadd{Exploiting the don't-care conditions in Equation~\eqref{eq:dc_care} and the candidate divisors}, 
\cdel{the SAT miter introduced previously confirms that the original pivot function $F=a\oplus d$ can be equivalently expressed using only these in-die divisors.}\cadd{we can use the method 
in Section~\ref{subsec:sat_resub} to construct \cdel{a new implementation}\cadd{an equivalent function} of $F$. First, the cross-die fanin $a$ is selected and excluded, while the remaining 
fanin $d$ is extracted as $BaseDivisors$. 
Following Algorithm~\ref{alg:algorithm2}, two types of divisor subsets are examined. 
The first consists only of $BaseDivisors$ (i.e., $S_d' = \{d\}$). 
The second augments $BaseDivisors$ with an additional in-die divisor $Y$, 
resulting in $S_d' = \{d, Y\}$. 
The SAT-based existence check returns true for $S_d' = \{d, Y\}$, 
indicating that $F$ can be equivalently expressed using only in-die divisors. 
The interpolated function is therefore derived as}
\begin{equation}
F' = Y \oplus d. \label{eq:f_new}
\end{equation}%

\cadd{As shown in Table~\ref{tab:truth_table},}
the \cdel{rewritten}\cadd{new} function $F'$ is equivalent to $F$ on all care minterms\cadd{,
which ensures the functional correctness after resubstitution.}
After resubstituting $F$ by $F'$, the post-resynthesis \cdel{circuit}\cadd{netlist} is shown in Figure~\ref{fig:resub_example_b}. $F'$ no longer depends on the \cadd{inter}-die signal $a$, while the necessary
\cadd{inter}-die dependency on $X$ is preserved via node $Y$. 
The amount of SLLs $(N_{sll})$ of this small example \cdel{decreased}\cadd{decreases} from 2 to 1.

\begin{table}[htbp]
\centering
\small
\caption{Truth table of the pivot node before and after resubstitution.
Gray rows correspond to \textbf{don't-care} conditions\cdel{ induced by the constraint $b=c$}\cadd{ ($b\neq c$)}.
Signals $Y$ and $d$ are in-die divisors used to reconstruct $F$.}
\label{tab:truth_table}
\begin{tabular}{cccc|c|c|c|c|c}
\hline
$a$ & $b$ & $c$ & $d$
& $X$
& $Y$
& \textbf{$F$ (orig.)}
& \textbf{$F'$ (new)}
& \textbf{care? }\\
\hline
0 & 0 & 0 & 0 & 0 & 0 & \textbf{0} & \textbf{0} & 1 \\
0 & 0 & 0 & 1 & 0 & 0 & \textbf{1} & \textbf{1} & 1 \\
\rowcolor{gray!20}
0 & 0 & 1 & 0 & 0 & 1 & 0 & 1 & 0 \\
\rowcolor{gray!20}
0 & 0 & 1 & 1 & 0 & 1 & 1 & 0 & 0 \\
\rowcolor{gray!20}
0 & 1 & 0 & 0 & 1 & 1 & 0 & 1 & 0 \\
\rowcolor{gray!20}
0 & 1 & 0 & 1 & 1 & 1 & 1 & 0 & 0 \\
0 & 1 & 1 & 0 & 1 & 0 & \textbf{0} & \textbf{0} & 1 \\
0 & 1 & 1 & 1 & 1 & 0 & \textbf{1} & \textbf{1} & 1 \\
1 & 0 & 0 & 0 & 1 & 1 & \textbf{1} & \textbf{1} & 1 \\
1 & 0 & 0 & 1 & 1 & 1 & \textbf{0} & \textbf{0} & 1 \\
\rowcolor{gray!20}
1 & 0 & 1 & 0 & 1 & 0 & 1 & 0 & 0 \\
\rowcolor{gray!20}
1 & 0 & 1 & 1 & 1 & 0 & 0 & 1 & 0 \\
\rowcolor{gray!20}
1 & 1 & 0 & 0 & 0 & 0 & 1 & 0 & 0 \\
\rowcolor{gray!20}
1 & 1 & 0 & 1 & 0 & 0 & 0 & 1 & 0 \\
1 & 1 & 1 & 0 & 0 & 1 & \textbf{1} & \textbf{1} & 1 \\
1 & 1 & 1 & 1 & 0 & 1 & \textbf{0} & \textbf{0} & 1 \\
\hline
\end{tabular}
\end{table}


\subsection{\xadd{Partition Imbalance Induced by Inter-Die Resynthesis}}\label{subsec:tradeoff}

In addition to minimizing the number of inter-die connections, 
partitioning-based \cdel{multi-die}\cadd{multi-die FPGA CAD} flows\cdel{must also}\cadd{ are also expected to} maintain a balanced distribution of logic resources across dies. 
Partition balance is a key feasibility constraint: a severely imbalanced die can become overutilized, leading to packing/placement failures or degraded QoR.
A commonly used metric is the partition imbalance ratio, denoted by $\rho$, which quantifies the deviation of logic distribution from an ideal uniform partition~\cite{raveena2022slip}. It is defined as
\begin{equation}
\label{eq:imbalance}
\rho = \max_{k \in \{0,\dots,K-1\}} 
\left( \frac{|V_k|}{|V|/K} \right),
\end{equation}%
where $K$ is the number of dies, $|V_k|$ denotes the number of logic nodes assigned to die $k$, and $|V|$ denotes the total number of logic nodes. In practice, an upper bound on $\rho$ called $UB$ is imposed during partitioning to ensure feasible packing and placement on each die.


\cdel{It is worth noting}\cadd{Note} that \cdel{interconnect-aware resubstitution for multi-die FPGAs}\cadd{the proposed \xadd{SLL-aware resynthesis}} does not strictly preserve the original partition balance. 
The success of resubstitution is inherently opportunistic and\deleted{depends on local structural conditions.}\xadd{ each success has an asymmetric and localized impact on the involved dies.}
\xadd{Figure~\ref{fig:resub_a} illustrates a representative scenario.} 
\xadd{On the destination die, the pivot node $v$ is replaced by another node $v'$, so the size typically remains unchanged.} \xadd{On the source die, the effect depends on the fanout structure. If the predecessor $v$ has no other fanouts, both $v$ and its MFFC become redundant and can be eliminated, thereby reducing the size of the corresponding die by one or more nodes. Otherwise, no node reduction occurs.}
%
This \xadd{size reduction} \xadd{behaviour} is \deleted{actually} the original purpose of \xadd{conventional}  \texttt{MFS}\xadd{ and is inherited by SLL-aware resynthesis, which consequently introduces uncertainty in die imbalance}.
As a result, the number of logic nodes assigned to individual dies may change, leading to a variation in the observed imbalance factor~$\rho$.
Since resubstitution is applied greedily in topological order, such node removals are \cdel{opportunity-driven and }difficult to predict analytically. Although the total number of LUTs in the netlist is guaranteed to be non-increasing (Line~\ref{alg:resub:check_area} in Algorithm~\ref{alg:md_resub_greedy}), the per-die logic distribution may vary slightly. Therefore, we report the imbalance factor before and after resynthesis in Section\cadd{s}~\ref{subsec:impact-connect}\cdel{,}\cadd{ and}~\ref{subsec:impact-placement} to evaluate its practical impact.

\section{Experimental Results}
\label{sec:exp}

This section evaluates the effectiveness of the proposed interconnect-aware
resynthesis\cadd{ method,} \texttt{ResynMD}, from both logic-level and physical
design perspectives.
We first assess the capability of \texttt{ResynMD} to reduce inter-die connectivity on purely combinational benchmarks, isolating the impact of logic resynthesis without interference from physical design effects.
We then study the sensitivity of the proposed approach to LUT size, which directly influences resubstitution opportunities and optimization effectiveness.
Finally, we evaluate the downstream impact of interconnect-aware resynthesis on partitioning-based multi-die placement, including wirelength, load balance, and runtime, using realistic heterogeneous FPGA benchmarks.
\subsection{Experimental Setup}
Our \cdel{SAT-based resubstitution resynthesis approach}\cadd{interconnect-aware logic resynthesis \xadd{method}\deleted{framework} for multi-die FPGAs, \texttt{ResynMD},} is implemented in C++ and Java. All experiments are conducted on a workstation equipped with an AMD Ryzen 9 7900X 12-core processor and 128~GB \cdel{of main memory}\cadd{RAM}, running a 64-bit Linux system.
For the logic synthesis and technology mapping stage in Figure~\ref{fig:flowchart},
we use \cadd{the} Yosys~\cite{yosys2025yosys} and \cdel{embedded}ABC~\cite{brayton2010abc} tools.
For the partitioning stage, we utilize an \texttt{hMETIS}-based~\cite{karypis1998hypergraph} timing-aware partitioning tool.
For the physical implementation stage,
we utilize the \deleted{standard} \texttt{AAPack}~\cite{Luu2011AAPack} clustering algorithm provided in \texttt{VTR8}~\cite{murray_vtr_2020},
and the \cdel{parallel }multi-die placement algorithm implemented in \texttt{LiquidMD}~\cite{raveena2024HEART, Dries2017LiquidFPT}.
No post-routing results are reported in this work, 
while our tool flow can be extended to evaluate routing performance in future work.

We fix the imbalance factor $UB$ of the partitioner to 1.25 across all experiments.
\xadd{We evaluate our approach using three benchmark suites, each targeting a distinct evaluation objective.}
\xadd{The EPFL~\cite{epfl} benchmark suite consists of purely combinational designs and is used to quantify the intrinsic effectiveness of the proposed logic-level resynthesis in reducing inter-die connections.
Since these circuits are relatively small and do not reflect realistic physical-design constraints, we report only connectivity-level metrics on EPFL \xadd{benchmarks}.}

\xadd{To evaluate physical-design impact in the presence of sequential logic, we additionally use a set of MCNC~\cite{yang1991logic} benchmarks.
These circuits contain flip-flops but no heterogeneous hard blocks, which enables isolating the effect of sequential elements while still allowing post-placement evaluation.}

\xadd{To evaluate physical-design impact on realistic heterogeneous FPGA designs, we utilize the Koios~\cite{arora2021koios} benchmark suite, which comprises designs featuring hard blocks such as DSPs and block RAMs.
In these circuits, hard macros and timing-optimized sequential elements are preserved and treated as structural boundaries, allowing us to assess whether SLL-aware resynthesis remains beneficial under constrained rewriting flexibility.
For both MCNC and Koios, we report post-placement results to quantify downstream physical-design effects.}


\deleted{To evaluate the effectiveness of our proposed \texttt{ResynMD} algorithm on large-scale circuits, we utilize the Koios benchmarks~\cite{arora2021koios}. To focus on the logic synthesis and re-substitution efficiency, we performed a logic-only transformation on the original heterogeneous designs. Specifically, all functional units including DSP macros and Block RAMs were decomposed into their equivalent soft-logic gate-level representations using Yosys\cadd{~\cite{yosys2025yosys}} and mapped into 4-input LUTs via ABC. This process ensures a homogeneous evaluation environment where the optimization impact on complex arithmetic logic can be measured without being obscured by fixed hard-block boundaries.}

\subsection{\xadd{Multi-Die Cost Modeling and Connectivity Metrics}}

During placement, wirelength is commonly estimated by summing the contributions of all nets, where the wirelength of each net is approximated using the half-perimeter wirelength (HPWL) of its bounding box. To account for net-specific characteristics, the HPWL of each net is multiplied by a weighting factor $q(n)$~\cite{Dries2017LiquidFPT,raveena2024HEART}. The overall wirelength cost is thus defined as
\begin{equation}
BBox_{cost}^{SD} = \sum_{n \in \text{nets}} q(n) \cdot \mathrm{HPWL}(n).
\label{eq:bbcost_sd}
\end{equation}
for a single-die FPGA.

\begin{figure}[!htbp]
     \centering
     \begin{subfigure}[b]{0.21\textwidth}
         \centering
         \includegraphics[width=\textwidth]{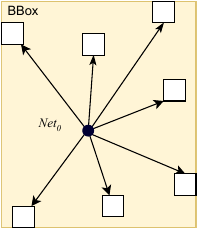}
         \caption{}
         \label{fig:bb_costa}
     \end{subfigure}
     \hfill
     \begin{subfigure}[b]{0.32\textwidth}
         \centering
         \includegraphics[width=\textwidth]{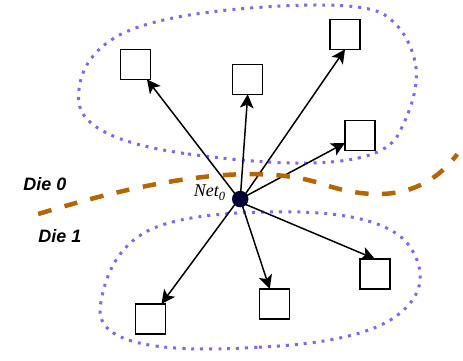}
         \caption{}
         \label{fig:bb_costb}
     \end{subfigure}
     \hfill
     \begin{subfigure}[b]{0.33\textwidth}
         \centering
         \includegraphics[width=\textwidth]{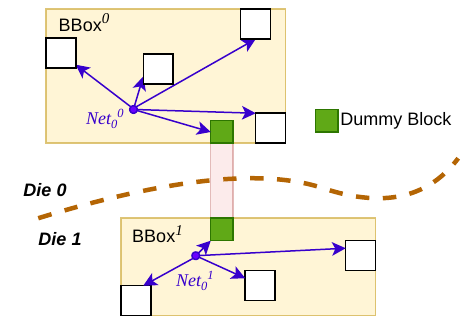}
         \caption{}
         \label{fig:bb_costc}
     \end{subfigure}
          \caption{Bounding-box (BB) wirelength estimation before and after die partitioning~\cite{raveena2022slip}.
(a) A non-partitioned net enclosed by a single bounding box within one die.
(b) After partitioning, the net spans multiple dies, forming die-local terminal clusters.
(c) The partitioned net is modeled using multiple die-local bounding boxes connected by inter-die segments, and the total BB wirelength accounts for both intra-die and inter-die components.}

        \label{fig:bb_cost}
        \Description{}
\end{figure}

For multi-die FPGA systems, \added{Figure~\ref{fig:bb_cost} illustrates how bounding-box wirelength estimation changes before and after die partitioning. For a non-partitioned design, all terminals of a net are enclosed by a single bounding box within one die (Figure~\ref{fig:bb_costa}). After partitioning, where partitions equate to dies—nets are classified as intra-die (terminals within one die) or inter-die (crossing boundaries). Intra-die nets use standard HPWL within the die region.
For inter-die nets, a single bounding box underestimates wirelength due to long interposer links. Thus, each net is decomposed into die-local segments (evaluated via HPWL with virtual boundary terminals) and an interposer segment modeled as a fixed-length SLL.}

\xadd{Accordingly, the total estimated wirelength of the multi-die system is computed as the sum of the per-die bounding-box costs defined in Equation~\eqref{eq:bbcost_sd}, together with the contribution of inter-die connections, and is expressed as}
\begin{equation}
\added{
\mathrm{BBox}_{\text{cost}}^{MD}
=
\sum_{d \in \mathcal{D}} BBox_{cost}^d +N_{sll} \cdot L_{sll},
}
\label{eq:bb_cost_system}
\end{equation}
\added{where $N_{sll}$ denotes the number of nets crossing the die boundary and $L_{\text{sll}}$ is the length of a single interposer link.
This modular formulation enables independent optimization of each die during placement while accurately accounting for the cost of timing-expensive inter-die connections.}

\subsubsection{Multi-Pin Inter-Die Nets and Edge-Level Objective}

For a multi-pin net whose sinks are distributed across different dies, current physical design flows typically introduce dummy blocks (e.g., SLL buffers) to model inter-die connections. These dummy blocks are placed at the source and sink sides of the SLLs, enabling consistent modeling in both tile-based and routing-based multi-die FPGA architectures. Furthermore, these \cadd{cross}-die sinks can share the same SLL channel, thereby reducing the second term of the system bounding-box cost in Equation~\ref{eq:bb_cost_system}.

\begin{figure}[!htbp]
    \centering
    \includegraphics[width=0.5\linewidth]{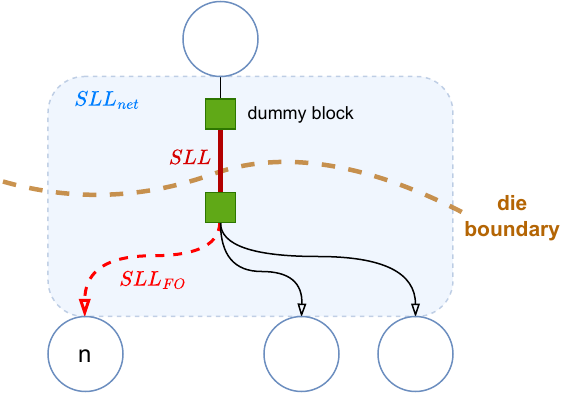}
    \caption{Illustration of a multi-pin inter-die net implemented via SLL dummy blocks. Eliminating the dependency on node $n$ removes one inter-die connection (i.e., reduces the \xadd{SLL net's} fanout by one), while the SLL net itself remains due to other cross-die sinks.}
    \label{fig:resub_net}
    \Description{}
\end{figure}
As illustrated in Figure~\ref{fig:resub_net}, in this example, an inter-die net may fan out to multiple sinks on the destination die through an SLL modeled as a dummy block. 
Each successful resubstitution in resynthesis is \emph{pivot-centric}: the optimization is performed with respect to a selected pivot node, and only its fanin structure is modified. Consequently,the resub removes at most the \cadd{inter}-die fanin edges incident to that pivot, while leaving the fanout structure of the source net unchanged. As illustrated in Figure~\ref{fig:resub_net}, when node $n$ is selected as the pivot, resubstitution can eliminate its dependency on the \cadd{inter}-die signal, other sinks of the same net remain unaffected. Therefore, a single resubstitution reduces exactly one SLL net's fanout ($SLL_{FO}$), but does not eliminate the entire SLL net if additional cross-die sinks still exist.
Therefore,  inter-die connectivity ($N_{sll}$) after physcial design implementation does not fully capture the structural impact of SLL-aware resynthesis.
For this reason, we refine the definition of inter-die connectivity, using two metrics: the net-level count $N_{\text{sll}}$ and the edge (connection)-level SLL fanout count $N_{\text{sll}}^{FO}$.

The $N_{\text{sll}}$ counts the number of SLL nets occupying SLL tracks after physical implementation, where multiple cross-die sinks may share the same SLL channel. In contrast, $N_{\text{sll}}^{FO}$ counts each \cadd{inter}-die fanout edge in the LUT\cadd{-level netlist} individually.
Each accepted resubstitution guarantees the removal of at least one inter-die edge, strictly reducing $N_{\text{sll}}^{FO}$ even if $N_{\text{sll}}$ remains unchanged.

\cadd{In what follows, we evaluate the effectiveness of \texttt{ResynMD} on inter-die connectivity in Section~\ref{subsec:impact-connect} and its impact on physical design in Section~\ref{subsec:impact-placement}.}

\subsection{Impact on Inter-Die Connectivity}\label{subsec:impact-connect}

\begin{figure}[!htbp]
    \centering

    \begin{subfigure}[t]{1\linewidth}
        \centering
        \includegraphics[
          width=\linewidth,
          height=0.11\textheight
        ]{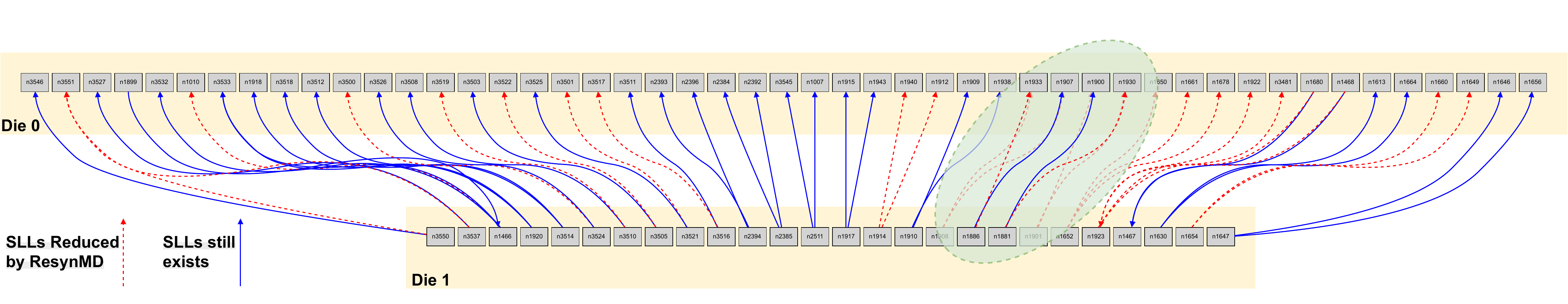}
        \caption{Inter-die connectivity.}
        \label{fig:resyn_example_a}
    \end{subfigure}

    \vspace{0.3em}

    \begin{subfigure}[t]{1\linewidth}
        \centering
        \includegraphics[
          width=.3\linewidth
        ]{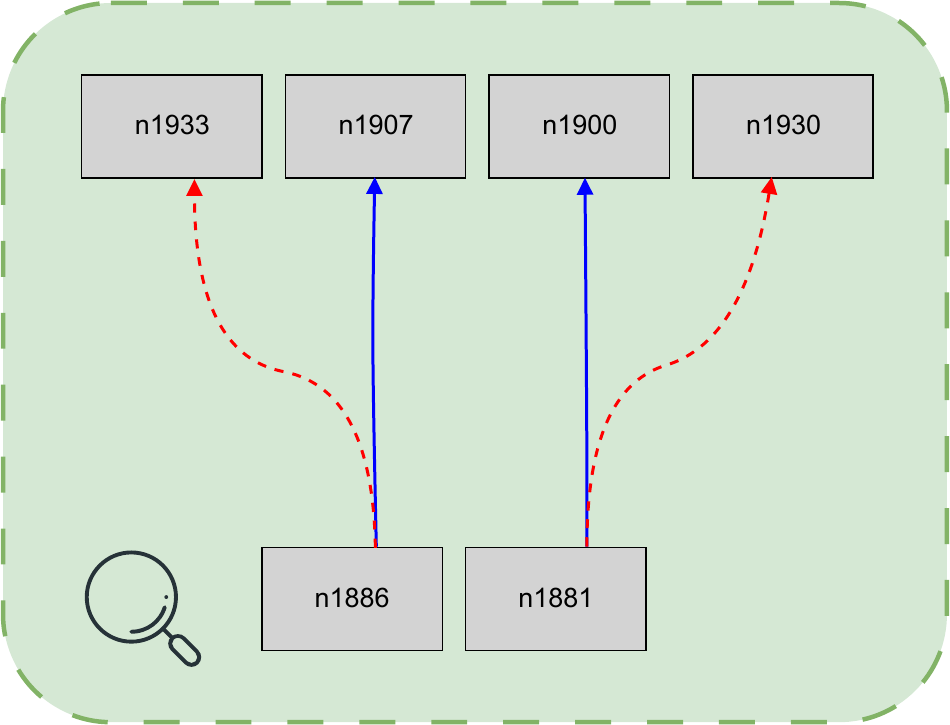}
        \caption{Partial optimized multi-pin inter-die net.}
        \label{fig:resyn_example_b}
    \end{subfigure}

    \caption{
    Inter-die connectivity of a two-die system for the $\textit{voter}$ benchmark from the EPFL suite.
    Logic blocks are grouped by die.
    Solid blue edges indicate inter-die connections that remain after resynthesis,
    while dashed red edges denote inter-die connections eliminated by interconnect-aware resynthesis.
    }
    \label{fig:resyn_example}
    \Description{}
\end{figure}

We evaluate the effectiveness of \texttt{ResynMD} on pure combinational logic benchmarks from the EPFL\cadd{ benchmark} suite.
The input netlists are first partitioned using \texttt{KaHyPar}~\cite{KaHyPar} with equal node weights, without considering timing criticality.
Subsequently, Algorithm~\ref{alg:md_resub_greedy} is applied to perform interconnect-aware logic resynthesis. As illustrated in Figure~\ref{fig:resyn_example}, the inter-die connectivity before and after \cdel{\texttt{ResynMD}}\cadd{interconnect-aware resynthesis} highlights the effectiveness of the proposed optimization, where plenty of \cadd{inter}-die connections (shown in red) are eliminated.

\cdel{Table~\ref{tab:sll_reduction} summarizes}\cadd{Tables~\ref{tab:sll_reduction_2die} and~\ref{tab:sll_reduction_3die} summarize} the impact of resynthesis on inter-die connectivity and partition balance under both 2-die and 3-die configurations.
Across the evaluated benchmarks, \texttt{ResynMD} reduces the number of inter-die connections in a substantial fraction of cases (11/20 for 2-die and 12/20 for 3-die), while leaving others unchanged, which reflects the circuit-dependent nature of resubstitution opportunities.


\begin{table}[!htbp]
\caption{Inter-die connections ($N_{\text{sll}}$) and imbalance factor $\rho$ before and after ResynMD on EPFL benchmarks (2-die configuration).}
\label{tab:sll_reduction_2die}
\centering
\small
\renewcommand{\arraystretch}{0.95}
\setlength{\tabcolsep}{1pt}

\begin{tabular}{l||cccc|cccc}
\hline
\multirow{2}{*}{\textbf{Bench.}}
& \multicolumn{4}{c|}{$N_{\text{sll}}$}
& \multicolumn{4}{c}{$\rho$} \\
\cline{2-9}
& Pre & Resyn & $\Delta$ & $\Delta(\%)$
& Pre & Resyn & $\Delta$ & $\Delta(\%)$ \\
\hline

cavlc
& 42 & 41 & -1 & -2.38
& 1.0070 & 1.0277 & +0.0207 & +2.05 \\

arbiter
& 5430 & 5171 & -259 & -4.77
& 1.0007 & 1.0000 & -0.0007 & -0.07 \\

voter
& 57 & 32 & -25 & -\textbf{43.86}
& 1.0380 & 1.0010 & -0.0370 & -3.57 \\

mem\_ctrl
& 1136 & 815 & -321 & -28.26
& 1.0147 & 1.0161 & +0.0014 & +0.14 \\

bar
& 526 & 458 & -68 & -12.93
& 1.0052 & 1.0084 & +0.0032 & +0.32 \\

sin
& 321 & 248 & -73 & -\textbf{22.74}
& 1.0161 & 1.0650 & +0.0489 & +4.81 \\

max
& 213 & 211 & -2 & -0.94
& 1.0374 & 1.0209 & -0.0165 & -1.59 \\

square
& 57 & 39 & -18 & -\textbf{31.58}
& 1.0194 & 1.0354 & +0.0160 & +1.56 \\

mult
& 1000 & 919 & -81 & -8.10
& 1.0003 & 1.0386 & +0.0383 & +3.83 \\

log2
& 3032 & 1554 & -1478 & -\textbf{48.74}
& 1.0328 & 1.1083 & +0.0755 & +7.31 \\

hyp
& 851 & 279 & -572 & -\textbf{67.21}
& 1.0252 & 1.0221 & -0.0031 & -0.30 \\

int2float
& 11 & 11 & 0 & 0
& 1.0698 & 1.0698 & 0 & 0 \\

\hline
\textbf{Avg.}
& -- & -- & \textbf{-263.45} & \textbf{-24.80}
& -- & -- & \textbf{+0.0133} & \textbf{+1.33} \\
\hline
\end{tabular}

\end{table}

\begin{table}[!htbp]
\caption{Inter-die connections ($N_{\text{sll}}$) and imbalance factor $\rho$ before and after ResynMD on EPFL benchmarks (3-die configuration).}
\label{tab:sll_reduction_3die}
\centering
\small
\renewcommand{\arraystretch}{0.95}
\setlength{\tabcolsep}{1pt}

\begin{tabular}{l||cccc|cccc}
\hline
\multirow{2}{*}{\textbf{Bench.}}
& \multicolumn{4}{c|}{$N_{\text{sll}}$}
& \multicolumn{4}{c}{$\rho$} \\
\cline{2-9}
& Pre & Resyn & $\Delta$ & $\Delta(\%)$
& Pre & Resyn & $\Delta$ & $\Delta(\%)$ \\
\hline

cavlc
& 61 & 56 & -5 & -8.20
& 1.0385 & 1.0316 & -0.0068 & -0.66 \\

arbiter
& 6322 & 6259 & -63 & -1.00
& 1.0042 & 1.0047 & +0.0005 & +0.05 \\

voter
& 81 & 41 & -40 & -\textbf{49.38}
& 1.0435 & 1.1038 & +0.0604 & +5.79 \\

mem\_ctrl
& 2233 & 1807 & -426 & -19.09
& 1.0770 & 1.0800 & +0.0030 & +0.28 \\

bar
& 816 & 671 & -145 & -17.77
& 1.0675 & 1.0967 & +0.0293 & +2.74 \\

sin
& 1412 & 1123 & -289 & -\textbf{20.48}
& 1.0426 & 1.2183 & +0.1758 & +16.86 \\

max
& 311 & 296 & -15 & -4.82
& 1.0187 & 1.0330 & +0.0143 & +1.40 \\

square
& 78 & 60 & -18 & -\textbf{23.08}
& 1.0345 & 1.0241 & -0.0104 & -1.00 \\

mult
& 1846 & 1704 & -142 & -7.69
& 1.0021 & 1.0721 & +0.0700 & +6.98 \\

log2
& 6930 & 3146 & -3784 & -\textbf{54.61}
& 1.0299 & 1.0591 & +0.0292 & +2.83 \\

hyp
& 1276 & 395 & -881 & -\textbf{69.04}
& 1.0253 & 1.0303 & +0.0050 & +0.49 \\

int2float
& 20 & 19 & -1 & -5.00
& 1.1163 & 1.1163 & 0.0000 & 0.00 \\

\hline
\textbf{Avg.}
& -- & -- & \textbf{-467.33} & \textbf{-27.38}
& -- & -- & \textbf{+0.0251} & \textbf{+2.35} \\
\hline
\end{tabular}

\vspace{1mm}
\end{table}

Benchmarks with large combinational cones and extensive logic sharing benefit the most, achieving reductions of up to \(67.21\%\) for 2-die ($69.04\%$ for 3-die) in inter-die connections.
On average, interconnect-aware resynthesis reduces inter-die connectivity by \(24.8\cadd{0}\%\) in the 2-die configuration and by \(\cdel{27.4}\cadd{27.38}\%\) in the 3-die configuration.
Even if the number of dies increases, each partition becomes smaller, which will reduce the availability of candidate divisors.
But at the same time, the total number of inter-die connections increases, creating more opportunities for eliminating logically unnecessary \cadd{inter}-die nets. As a result, the net reduction in $N_{\text{sll}}$ is slightly larger in the 3-die case.

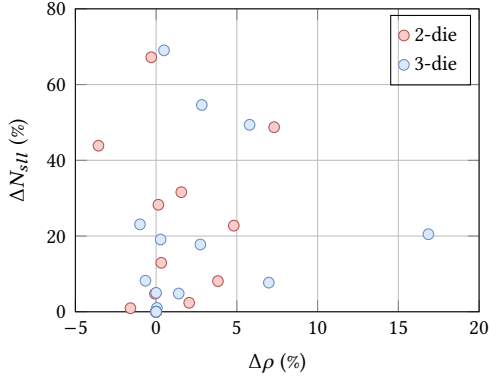
\begin{figure}[!htbp]
\centering
\begin{tikzpicture}
\begin{axis}[
    width=0.5\linewidth,
    height=0.4\linewidth,
    xlabel={$\Delta\rho$ (\%)},
    ylabel={$\Delta N_{sll}$ (\%)},
    xmin=-5, xmax=20,
    ymin=0, ymax=80,
    grid=both,
    tick label style={font=\footnotesize},
    label style={font=\footnotesize},
    legend style={
        font=\footnotesize,
        at={(0.98,0.98)},
        anchor=north east,
        draw=black,
        fill=none,
    }
]

\addplot[
    only marks,
    mark=*,
    dieTwol,
    fill=dieTwof,
    fill opacity=1,
    draw opacity=1,
] coordinates {
    ( 2.05,  2.38)
    (-0.07,  4.77)
    (-3.57, 43.86)
    ( 0.14, 28.26)
    ( 0.32, 12.93)
    ( 4.81, 22.74)
    (-1.59,  0.94)
    ( 1.56, 31.58)
    ( 3.83,  8.10)
    ( 7.31, 48.74)
    (-0.30, 67.21)
    ( 0, 0)
    ( 0, 0 )
    ( 0, 0 )
    ( 0, 0 )
    ( 0, 0 )
    ( 0, 0 )
    ( 0, 0 )
    ( 0, 0 )
    ( 0, 0 )
};
\addlegendentry{2-die}

\addplot[
    only marks,
    mark=*,
    dieThreel,
    fill=dieThreef,
    fill opacity=1,
    draw opacity=1,
] coordinates {
    (-0.66,  8.20)
    ( 0.05,  1.00)
    ( 5.79, 49.38)
    ( 0.28, 19.09)
    ( 2.74, 17.77)
    (16.86, 20.48)
    ( 1.40,  4.82)
    (-1.00, 23.08)
    ( 6.98,  7.69)
    ( 2.83, 54.61)
    ( 0.49, 69.04)
    ( 0, 5 )
    ( 0, 0 )
    ( 0, 0 )
    ( 0, 0 )
    ( 0, 0 )
    ( 0, 0 )
    ( 0, 0 )
    ( 0, 0 )
    ( 0, 0 )
};
\addlegendentry{3-die}



\end{axis}
\end{tikzpicture}
\caption{Trade-off between inter-die connectivity reduction and imbalance variation under 2-die and 3-die configurations.}

\label{fig:scatter_tradeoff}
\Description{}
\end{figure}

\deleted{This} \xadd{The} increased effectiveness is \xadd{likely }accompanied by a larger variation in partition imbalance.
Compared to the 2-die case, the 3-die configuration exhibits a higher average increase in imbalance ratio calculated from Equation~\ref{eq:imbalance} (\(2.35\%\) versus \(1.33\%\)), reflecting the more extensive structural changes introduced by resynthesis. Nevertheless, all observed imbalance values remain within practical bounds, and stricter imbalance constraints can be imposed during the initial partitioning stage if needed.
Figure~\ref{fig:scatter_tradeoff} further illustrates the relationship between inter-die connectivity reduction and imbalance variation. No strong correlation is observed between $\Delta\rho$ and $\Delta N_{sll}$, and several circuits achieve substantial SLL reduction with negligible imbalance change.
This suggests that the effectiveness of \texttt{ResynMD} does not inherently rely on increasing partition imbalance.




\subsubsection{\added{Physical Implications of Edge-Level SLL Reduction}}


Although \texttt{ResynMD} does not always reduce the number of SLL nets ($N_{\text{sll}}$), it consistently reduces the number of inter-die connections ($N_{\text{sll}}^{FO}$). This distinction is important from a physical design perspective. As illustrated by the green zoomed-in nets in Figure~\ref{fig:resyn_example_b}, one inter-die connection is removed, while another remains.
High-fanout inter-die nets impose disproportionate routing and timing pressure, as they load inter-die resources and near-die boundary routing resources, and increase routing complexity within the destination die.

Reducing the number of inter-die SLL fanouts ($N_{\text{sll}}^{FO}$) decreases SLL loading and mitigates routing congestion around SLL endpoints. Since SLL sink locations are fixed at the die boundary and already incur substantial inter-die delay, signals traversing SLLs typically exhibit high arrival times and timing criticality.
Under congestion, routing detours near these fixed interface points can further increase delay and exacerbate timing violations. By reducing \cadd{inter}-die fanout, the proposed method alleviates local congestion pressure around SLL interfaces, thereby improving routability and timing robustness even when the total number of occupied SLL tracks remains unchanged.

\subsection{Impact of LUT Size on SLL-Aware Resynthesis Effectiveness}
The effectiveness of SLL-aware resynthesis is strongly influenced by the target LUT size, which directly determines structural flexibility and divisor availability in the mapped \cdel{network}\cadd{LUT-level netlist}. Smaller LUT sizes produce finer-grained logic decomposition, resulting in a larger number of nodes with simpler Boolean functions. This increases the number of candidate divisors within each resubstitution window and reduces SAT constraint complexity, thereby improving the likelihood of finding valid in-die replacements.
In contrast, larger LUT sizes consolidate logic into fewer nodes with higher functional complexity. 
This reduces the available divisor set and enlarges the SAT search space, making it more difficult to identify feasible substitutions that reduce inter-die connectivity.

\xadd{To quantify this sensitivity, we map the EPFL benchmarks to LUT-4/5/6 \deleted{, LUT5, and LUT6 architectures} and measure the reduction in inter-die SLL nets after resubstitution, as summarized in Table~\ref{tab:lut_size_sensitivity}. In this experiment, we use a larger window size than in the previous effectiveness study to avoid overly restricting the divisor space. Unlike the previous experiments that use min-cut partitioners (\texttt{KaHyPar}/\texttt{hMetis}), the initial die assignment here is generated by a deterministic hash-based labeling scheme to eliminate partitioning variability across different LUT mappings. Since this pseudo-partition is not optimized for cut minimization, the initial inter-die connectivity is typically higher, providing more opportunities for resubstitution to remove cross-die nets. As a result, the absolute SLL reductions are larger than those in Section~\ref{subsec:impact-connect}, but still adequate for comparing LUT-size impact.

On average, LUT4 achieves the largest reduction (2243.42 nets), followed by LUT5 (1838.26), while LUT6 shows the smallest improvement (1622.11). 
This trend is consistent across most benchmarks, indicating that LUT granularity plays a critical role in interconnect-aware optimization. Post-partition resubstitution benefits from finer-grained logic representations, while larger LUT sizes limit optimization opportunities due to reduced structural flexibility.}

\begin{table}[!htbp]
\centering
\small
\caption{Reduction in inter-die SLL edges $\Delta$$N_{sll}$ under different LUT sizes (LUT-4/5/6).}
\label{tab:lut_size_sensitivity}
\setlength{\tabcolsep}{4pt}

\begin{tabular}{l|ccc||l|ccc}
\hline
\textbf{Bench.} & LUT4 & LUT5 & LUT6 
& \textbf{Bench.} & LUT4 & LUT5 & LUT6 \\
\hline
int2float & 46 & 33 & 25 
& sin & 1382 & 1259 & 1155 \\

ctrl & 23 & 6 & 5
& max & 568 & 477 & 388 \\

router & 37 & 30 & 30
& square & 3789 & 3351 & 2949 \\

cavlc & 179 & 115 & 53
& sqrt & 3255 & 3205 & 2660 \\

priority & 145 & 113 & 105
& multiplier & 5107 & 4795 & 4328 \\

dec & 32 & 31 & 31
& log2 & 6889 & 6141 & 5610 \\

i2c & 177 & 129 & 104
& div & 4538 & 3290 & 4032 \\

arbiter & 2215 & 1729 & 1417
& voter & 1960 & 1772 & 1425 \\

mem\_ctrl & 9361 & 7374 & 6077
& & & &\\
\hline

\textbf{Total} & \textbf{42625} & 34927 & 30820 
&
\textbf{Average} & \textbf{2243.42} & 1838.26 & 1622.11 
\\



\hline
\end{tabular}
\end{table}


\subsection{Impact on Physical Design (Placement)}\label{subsec:impact-placement}
The target chip is configured as a 325 $\times$ 436–tile FPGA, partitioned into a 2$\times$1 multi-die layout. Each die spans 325 × 218 tiles. The two dies are interconnected through 36 rows of super-long lines, each modeled with a fixed delay of 2.224 \textit{ns}. \xadd{We first evaluate the effectiveness of the proposed method on the MCNC benchmarks and assess its impact on physical design using sequential circuits. We then extend the study to designs containing hard blocks from the Koios benchmark suite.}
\begin{sidewaystable}[!htbp]

\centering
\footnotesize

\caption{Impact of \texttt{ResynMD} on inter-die connectivity, placement quality, load balance, and placement runtime across MCNC benchmarks.}

\label{tab:unified_metrics_A}
\setlength{\tabcolsep}{2pt}

\begin{tabular}{l||cccc||cccc||cccc||cccc||cccc||cccc}
\hline
\multirow{2}{*}{\textbf{Bench.}}
& \multicolumn{4}{c||}{$N_{\text{sll}}$}
& \multicolumn{4}{c||}{$N_{\text{sll}}^{FO}$}
& \multicolumn{4}{c||}{$BBox_{cost}^{2die}$}
& \multicolumn{4}{c||}{$\rho_{pre}$}
& \multicolumn{4}{c||}{$\rho_{post}$}
& \multicolumn{4}{c}{$\text{Runtime}_{place}$} \\
\cline{2-25}
& P & R & $\Delta$ & $\%$
& P & R & $\Delta$ & $\%$
& P & R & $\Delta$ & $\%$
& P & R & $\Delta$ & $\%$
& P & R & $\Delta$ & $\%$
& P & R & $\Delta$ & $\%$ \\
\hline


apex2
& 66 & 64 & -2 & -3.03
& 400 & 293 & -107 & -26.75
& 582.9 & 575.3 & -7.6 & -1.30
& 1.130 & 1.129 & -0.0005 & -0.04
& 1.128 & 1.123 & -0.005 & -0.41
& 0.580 & 0.564 & -0.016 & -2.73 \\

apex4
& 134 & 134 & 0 & 0.00
& 872 & 827 & -45 & -5.16
& 601.8 & 594.5 & -7.3 & -1.21
& 1.151 & 1.146 & -0.005 & -0.41
& 1.131 & 1.138 & +0.007 & +0.60
& 0.517 & 0.522 & +0.005 & +0.99 \\

bigkey
& 6 & 6 & 0 & 0.00
& 784 & 784 & 0 & 0.00
& 459.9 & 424.8 & -35.1 & -7.64
& 1.006 & 1.006 & 0.000 & 0.00
& 1.000 & 1.000 & 0.000 & 0.00
& 0.644 & 0.651 & +0.006 & +1.01 \\

clma
& 143 & 141 & -2 & -1.40
& 3821 & 2879 & -942 & -24.65
& 1996.4 & 2017.8 & +21.4 & +1.07
& 1.142 & 1.135 & -0.007 & -0.59
& 1.141 & 1.142 & +0.002 & +0.14
& 0.791 & 0.791 & 0.000 & -0.01 \\

des
& 11 & 11 & 0 & 0.00
& 581 & 581 & 0 & 0.00
& 325.0 & 346.3 & +21.3 & +6.56
& 1.026 & 1.026 & 0.000 & 0.00
& 1.025 & 1.024 & -0.0002 & -0.02
& 0.723 & 0.706 & -0.018 & -2.45 \\

diffeq
& 49 & 49 & 0 & 0.00
& 995 & 980 & -15 & -1.51
& 468.0 & 465.8 & -2.2 & -0.47
& 1.130 & 1.130 & 0.000 & 0.00
& 1.128 & 1.134 & +0.006 & +0.53
& 0.545 & 0.539 & -0.006 & -1.17 \\

dsip
& 3 & 3 & 0 & 0.00
& 560 & 560 & 0 & 0.00
& 281.0 & 275.5 & -5.4 & -1.94
& 1.008 & 1.008 & 0.000 & 0.00
& 1.006 & 1.006 & 0.000 & 0.00
& 0.624 & 0.627 & +0.003 & +0.52 \\



ex5p
& 186 & 165 & -21 & -11.29
& 922 & 679 & -243 & -26.36
& 758.3 & 673.8 & -84.5 & -11.14
& 1.167 & 1.145 & -0.022 & -1.90
& 1.155 & 1.126 & -0.029 & -2.51
& 0.559 & 0.538 & -0.021 & -3.82 \\

frisc
& 173 & 173 & 0 & 0.00
& 3075 & 2772 & -303 & -9.85
& 1142.8 & 1127.8 & -15.0 & -1.31
& 1.064 & 1.064 & 0.000 & 0.00
& 1.054 & 1.052 & -0.002 & -0.18
& 0.639 & 0.633 & -0.006 & -0.94 \\

misex3
& 117 & 116 & -1 & -0.85
& 806 & 693 & -113 & -14.02
& 596.3 & 595.9 & -0.4 & -0.07
& 1.157 & 1.144 & -0.013 & -1.11
& 1.148 & 1.149 & +0.002 & +0.14
& 0.536 & 0.533 & -0.003 & -0.60 \\

pdc
& 312 & 303 & -9 & -2.88
& 5577 & 4547 & -1030 & -18.47
& 1469.9 & 1411.7 & -58.2 & -3.96
& 1.171 & 1.155 & -0.015 & -1.31
& 1.152 & 1.186 & +0.033 & +2.89
& 0.644 & 0.661 & +0.017 & +2.61 \\

s38417
& 46 & 46 & 0 & 0.00
& 2276 & 2204 & -72 & -3.16
& 1047.1 & 1070.6 & +23.5 & +2.25
& 1.001 & 1.001 & 0.000 & 0.00
& 1.000 & 1.003 & +0.003 & +0.26
& 0.724 & 0.717 & -0.007 & -1.03 \\

s38584.1
& 21 & 21 & 0 & 0.00
& 479 & 471 & -8 & -1.67
& 1110.2 & 1158.2 & +48.1 & +4.33
& 1.106 & 1.106 & 0.000 & 0.00
& 1.082 & 1.088 & +0.006 & +0.56
& 0.737 & 0.753 & +0.016 & +2.22 \\

seq
& 112 & 111 & -1 & -0.89
& 671 & 560 & -111 & -16.54
& 732.0 & 768.8 & +36.8 & +5.03
& 1.130 & 1.125 & -0.005 & -0.43
& 1.126 & 1.128 & +0.002 & +0.18
& 0.596 & 0.599 & +0.003 & +0.42 \\

tseng
& 35 & 35 & 0 & 0.00
& 636 & 628 & -8 & -1.26
& 478.0 & 472.2 & -5.9 & -1.23
& 1.150 & 1.151 & +0.001 & +0.07
& 1.072 & 1.107 & +0.035 & +3.31
& 0.590 & 0.577 & -0.013 & -2.12 \\

\hline
\textbf{Avg.}
& -- & -- & \textbf{-2.40} & \textbf{-1.65}
& -- & -- & \textbf{-199.7} & \textbf{-12.29}
& -- & -- & \textbf{-4.69} & \textbf{-1.13}
& -- & -- & \textbf{-0.0044} & \textbf{-0.40}
& -- & -- & \textbf{+0.0040} & \textbf{+0.37}
& -- & -- & \textbf{-0.0027} & \textbf{-0.44} \\
\hline
\end{tabular}
\end{sidewaystable}

\begin{table*}[!htbp]
\centering
\small
\caption{Impact of ResynMD on timing cost (total arrival time) and critical-path delay (CPD) across MCNC benchmarks, both reported in \textit{ns}.}
\label{tab:unified_metrics_B}
\setlength{\tabcolsep}{3pt}

\begin{tabular}{l||cccc||cccc}
\hline
\multirow{2}{*}{\textbf{Bench.}}
& \multicolumn{4}{c||}{Timing cost (\textit{ns})}
& \multicolumn{4}{c}{CPD (\textit{ns})} \\
\cline{2-9}
& Prev & Resysn & $\Delta$ & $\%$
& Prev & Resysn & $\Delta$ & $\%$ \\
\hline


apex2
& 6267.0 & 6094.5 & -173.0 & -2.75
& 42.71 & 31.16 & -11.56 & -27.06 \\

apex4
& 2928.4 & 2768.9 & -159.0 & -5.45
& 42.25 & 38.42 & -3.83 & -9.07 \\

bigkey
& 4285.2 & 4045.5 & -240.0 & -5.59
& 33.75 & 33.98 & +0.23 & +0.68 \\

clma
& 9243.5 & 10042.0 & +799.0 & +8.64
& 34.64 & 39.41 & +4.77 & +13.77 \\

des
& 2896.7 & 3112.3 & +216.0 & +7.45
& 36.37 & 34.69 & -1.67 & -4.60 \\

diffeq
& 1356.4 & 1183.9 & -173.0 & -12.72
& 38.24 & 31.10 & -7.14 & -18.68 \\

dsip
& 2406.0 & 2386.3 & -19.7 & -0.82
& 31.40 & 27.27 & -4.13 & -13.15 \\



ex5p
& 3002.3 & 2553.5 & -449.0 & -14.95
& 46.00 & 39.55 & -6.44 & -14.01 \\

frisc
& 4429.4 & 4210.9 & -219.0 & -4.93
& 33.93 & 45.84 & +11.92 & +35.12 \\

misex3
& 5070.9 & 5662.7 & +592.0 & +11.67
& 36.25 & 29.44 & -6.81 & -18.78 \\

pdc
& 7935.8 & 7214.6 & -721.0 & -9.09
& 43.67 & 44.07 & +0.41 & +0.93 \\

s38417
& 2872.2 & 2954.1 & +81.9 & +2.85
& 29.21 & 24.77 & -4.43 & -15.18 \\

s38584.1
& 2018.8 & 1852.1 & -167.0 & -8.25
& 30.74 & 30.54 & -0.20 & -0.64 \\

seq
& 5001.5 & 6685.2 & +1680.0 & +33.66
& 34.36 & 38.20 & +3.83 & +11.15 \\

tseng
& 918.5 & 921.8 & +3.3 & +0.36
& 28.04 & 26.98 & -1.06 & -3.76 \\

\hline
\textbf{Avg.}
& -- & -- & \textbf{+70.10} & \textbf{+0.01}
& -- & -- & \textbf{-1.74} & \textbf{-4.22} \\
\hline
\end{tabular}
\end{table*}

\begin{table*}[!htbp]
\caption{
Impact of \texttt{ResynMD} on physical design across Koios benchmarks
with reduced inter-die connectivity (2-die configuration).
}
\label{tab:koios}
{\centering
\small
\setlength{\tabcolsep}{3.2pt}
\begin{tabular}{l||cc|cccc}
\hline
\multirow{2}{*}{\textbf{Bench.}} &
\multicolumn{2}{c|}{Inter-die connectivity} &
\multicolumn{4}{c}{Post-placement metrics} \\
\cline{2-7}
 & $\Delta$ $N_{\text{sll}}$ 
 & $\Delta$ $N_{\text{sll}}^{FO}$  &
 $\Delta$ $BBox_{cost}^{2die}$ (\%) &
 $\Delta$ Timing (\%) &
 $\Delta$ CPD (\%) &
 $\Delta$ Runtime (\%) \\
\hline
robot\_rl
 & 0 (0.00\%) 
 & -189 (-10.79\%) 
 & -2.18 
 & \textbf{+5.77 $\uparrow$}
 & +0.42 
 & -0.88 \\

tpu\_like.medium
 & -2 (-0.61\%) 
 & -980 (-2.68\%) 
 & \textbf{-7.25 $\downarrow$}
 & +3.85 
 & \textbf{+5.80 $\uparrow$}
 & +0.74 \\

lstm
 & 0 (0.00\%) 
 & -145 (-0.25\%) 
 & \textbf{-8.89 $\downarrow$}
 & \textbf{-5.50 $\downarrow$}
 & -2.71 
 & -1.35 \\

reduction\_layer
 & 0 (0.00\%) 
 & -8 (-0.10\%) 
 & -0.17 
 & +4.86 
 & \textbf{-20.23 $\downarrow$}
 & -2.04 \\

softmax
 & 0 (0.00\%) 
 & -4 (-0.24\%) 
 & -2.68 
 & +1.02 
 & \textbf{-21.42 $\downarrow$}
 & -1.22 \\

dla\_like.medium
 & 0 (0.00\%) 
 & -1 (-0.00\%) 
 & +1.80 
 & \textbf{+7.15 $\uparrow$}
 & -1.08 
 & -2.69 \\

dla\_like.small
 & 0 (0.00\%) 
 & -7 (-0.01\%) 
 & +0.13 
 & \textbf{+21.92 $\uparrow$}
 & -1.54 
 & -2.70 \\
\hline
\end{tabular}

}\vspace{2mm}
\raggedright
\footnotesize
*Arrows indicate significant variations ($|\Delta| > 5\%$).
\end{table*}

\subsubsection{Inter-Die Connectivity Reduction}

Table~\ref{tab:unified_metrics_A} reports the impact of \texttt{ResynMD} on inter-die connectivity under the 2-die configuration. Across the MCNC~\cite{yang1991logic} benchmarks, the proposed method reduces \cadd{inter}-die fanout edges by 12.29\% on average, while the net-level reduction is 1.65\%.
The discrepancy between edge-level and net-level improvements is expected. Multiple cross-die sinks may share a single SLL track, and removing individual fanout edges does not necessarily eliminate the occupied SLL channel. Nevertheless, reducing \cadd{inter}-die fanout edges lowers SLL loading and simplifies inter-die connectivity at the structural level.

\subsubsection{Placement Quality and Load Balance}

The reduction in \cadd{inter}-die connectivity translates to modest improvements in placement quality.
On MCNC benchmarks, the average bounding-box cost decreases by 1.13\%. Although the average BBox reduction appears modest, it reflects a more spatially compact placement solution. Reducing \cadd{inter}-die fanout edges shortens the effective span of inter-die nets, which decreases routing demand and alleviates congestion in multi-die architectures.
Circuits exhibiting larger edge-level reductions (e.g., \textit{ex5p} and \textit{pdc}) show more noticeable BBox improvements, suggesting that reduced inter-die coupling can alleviate global placement pressure. \xadd{These effects are expect to become more pronounced during routing, as reduced congestion could potentially lower the number of rip-up-and-reroute iterations. A detailed routing-level analysis is left for future work.}

Die-level load balance remains largely stable.
The pre-pack imbalance factor changes by $-0.40$\% on average, while the post-pack imbalance varies by $+0.37\%$. These small variations indicate that interconnect-aware resynthesis does not significantly distort partition balance and remains compatible with partitioning-based multi-die flows. The imbalance change at the LUT-level
is smaller than the intrinsic variation introduced by the packing stage.
Placement runtime is also stable, with an average change of $-0.44$\%, demonstrating that the proposed method does not introduce additional overhead in the physical design stage.

\xadd{In cases of \deleted{severe} partition imbalance, the optimization can be restricted to nodes in the smaller die subgraph, potentially reducing the effective size of the larger die and improving partition balance.}
\xadd{Since the optimization objective adheres to the area constraint defined in Equation~\ref{eq:obj_target}, 
the total circuit size is guaranteed to be non-increasing, with only a possible increase in edge count due to additional fanout connections of selected divisors. Therefore, even if the imbalance factor slightly increases (e.g., due to further reduction of the smaller die), the downstream physical design stages are not adversely affected compared with the baseline flow.}
\subsubsection{Timing Behavior}

The impact on timing is circuit-dependent, as shown in Table~\ref{tab:unified_metrics_B}.
While several benchmarks (e.g., \textit{ex5p}, \textit{diffeq}, and \textit{pdc}) exhibit timing improvement, some of the others experience degradation.
On average, the critical-path delay (CPD) decreases by $-4.22\%$. This non-uniform behavior is attributed to the absence of explicit timing constraints in the resynthesis objective. While minimizing \cadd{inter}-die fanout effectively alleviates routing congestion and SLL overhead, logic resubstitution carries the risk of selecting timing-critical signals as divisors for node interpolation. \xadd{The example in Figure~\ref{fig:resub_example_b} illustrates this issue, although the SLL at the pivot node is eliminated, the selected divisor $Y$ lies on a timing-critical path, and its own fanin already crosses die boundaries.} This process inadvertently extends the critical path and sabotages timing slack. Therefore, inter-die connectivity reduction alone does not guarantee timing improvement. Incorporating slack-based information as the cost function of SLL-aware logic resynthesis can prevent timing degradation; however, this con will comes at the expense of resubstitution efficacy.
These observations highlight the complex interaction between logic-level restructuring and downstream multi-die placement optimization.

\subsubsection{Heterogeneous Koios Benchmarks}\label{subsect:Koios}

As shown in Table~\ref{tab:koios}, in Koios benchmarks that contain hard blocks, net-level SLL reductions $\Delta N_{sll}$ are limited, and edge-level improvements $\Delta N_{sll}^{FO}$ are generally small. This is expected, as inter-die connectivity in heterogeneous designs is often dominated by hard macros that cannot be modified by LUT-level resubstitution. 
In the current Koios flow, the circuits are technology-mapped and timing-optimized prior to resynthesis. Hard blocks and sequential latches are treated as black boxes, and their I/Os are exposed to the LUT\cadd{-level netlist} as primary inputs and outputs. As a result, logical continuity across these modules is structurally interrupted. During window-based resubstitution, such boundaries restrict the availability of valid divisors, since signals cannot propagate across black-box interfaces. \deleted{This significantly limits rewriting flexibility and reduces the opportunity to remove \cadd{inter}-die connections.}

Nevertheless, \xadd{due to computational resource constraints, we conducted full physical design experiments on 15 Koios circuits, among which 8 exhibit measurable reductions in inter-die connectivity under \texttt{ResynMD}. }\deleted{notably,}certain designs such as \textit{lstm} and \textit{tpu\_like.medium} exhibit BBox reductions of up to 8.89\% and 7.25\% respectively, with an average of 2.75\% reduction on 7 listed circuits. Although net-level SLL reduction is limited in heterogeneous circuits dominated by hard macros, these results indicate that even modest edge-level improvements within combinational regions can reshape spatial placement organization. In terms of timing, the average CPD across the seven listed Koios benchmarks decreases by 5.82\%, although the behavior remains design-dependent. Significant improvements are observed in designs such as \textit{reduction\_layer} and \textit{softmax}, where CPD decreases by over 20\%. These results suggest that even when net-level SLL reduction is structurally constrained, localized edge-level simplification can still mitigate \cadd{inter}-die routing overhead and reduce critical-path sensitivity to SLL delay.

These observations imply that SLL-aware resynthesis would \deleted{likely} be more effective if applied before \deleted{timing-driven mapping and }hard-block binding \xadd{and physical synthesis}. Performing logic restructuring earlier in the flow would provide greater structural freedom and potentially expose more opportunities for reducing inter-die connectivity.

\section{Conclusion and Future Work}\label{sec:con}

\xadd{This paper presents \texttt{ResynMD}, an interconnect-aware logic resynthesis method integrated into a partitioning-based multi-die tool flow to reduce inter-die connections.}
To the best of our knowledge, this is the first work to \cdel{integrate LUT-network}\cadd{incorporate} logic resynthesis into a \cdel{2.5D/}multi-die FPGA CAD flow\cadd{ for inter-die connection optimization}.
By targeting LUT-level \cdel{combinational }logic \cdel{dependencies that dominate inter-die communication in partitioned designs}\cadd{that contributes substantially to inter-die connections}, 
\cdel{the proposed method complements}\cadd{\texttt{ResynMD} is orthogonal to} existing \xadd{partitioning-based}\deleted{partition-driven} flows and demonstrates that logic\cdel{-level restructuring is a practical and effective means to mitigate}\cadd{ resynthesis can reduce} inter-die \cdel{connectivity}\cadd{connections} that \cdel{cannot be eliminated by partitioning alone}\cadd{partitioning alone cannot eliminate}. 
Experimental results show that \texttt{ResynMD} \cdel{achieves substantial reductions in}\cadd{reduces} inter-die connections across a wide range of benchmarks, 
\cadd{with }particularly\cadd{ strong benefits} \cdel{on}\cadd{for} combinational \cdel{circuits }and soft-logic-dominated designs. 
\cadd{These improvements are achieved} while maintaining acceptable partition balance and 
\cdel{physical feasibility}\cadd{successful physical implementation}. 
\cdel{And these reductions translate into}\cadd{The reduction in inter-die connections also leads to}\cdel{measurable improvements in downstream placement quality}\cadd{ downstream effects on physical metrics after placement, such as wirelength and critical path delay}.

\cadd{Future work will incorporate routing into both the evaluation and the resynthesis objective,
enabling a more realistic assessment of inter-die routing resource usage and overall implementation quality.
Finally, although this work focuses on}\cdel{Beyond} LUT-based FPGA fabrics, 
the principle of interconnect-aware logic \cdel{restructuring}\cadd{resynthesis} can be extended to \cdel{ASIC }chiplet\cadd{-based ASIC systems} or 3D ICs, 
where \cadd{inter}-die\cadd{ or inter-layer} interconnects similarly incur high delay and routing cost. 
In such settings, the optimization would \cdel{take the form of}\cadd{operate on} technology-mapped \cdel{structural rewriting under  standard-cell constraints}\cadd{gate-level netlist} rather than \cdel{LUT-level resubstitution}\cadd{on LUT-level netlists,
by performing functionality-preserving transformations to reduce inter-die or inter-layer connections}.

\newpage

\bibliographystyle{unsrt} 

\end{document}